\documentclass[aps,prd,reprint,superscriptaddress,nofootinbib,floatfix,longbibliography]{revtex4-1}
\usepackage[dvipsnames]{xcolor}
\usepackage[utf8]{inputenc}
\usepackage[T1]{fontenc}
\usepackage{amssymb,amsmath,amsfonts,bm,slashed,soul,ragged2e,graphicx,epstopdf,hyperref,array,bm}
\usepackage[utf8]{inputenc}
\usepackage{colortbl,color,caption,subcaption}
\enlargethispage{\baselineskip}
\definecolor{steelblue}{RGB}{25,25,112}
\definecolor{dullblue}{rgb}{0,0.298,0.49}
\definecolor{darkred}{rgb}{0.545,0,0}
\definecolor{darkorange}{RGB}{222,132,69}
\definecolor{darkgreen}{RGB}{126,171,85}
\definecolor{blue2}{cmyk}{1, 0.1, 0.1, 0}
\hypersetup{colorlinks,linkcolor={darkred},citecolor={blue},urlcolor={dullblue}}
\DeclareCaptionJustification{justified}{\justifying}
\everymath{\displaystyle}
\usepackage{placeins}

\usepackage{comment}
\usepackage{braket}
\usepackage{amsmath}
\usepackage{soul}
\usepackage{makecell}
\usepackage{booktabs}
\usepackage{siunitx}
\usepackage{multirow}

\newcommand{\beq}{\begin{equation}}
\newcommand{\eeq}{\end{equation}}
\newcommand{\bea}{\begin{eqnarray}}
\newcommand{\eea}{\end{eqnarray}}

\newcommand{\gsim}{\lower.7ex\hbox{$\;\stackrel{\textstyle>}{\sim}\;$}}
\newcommand{\lsim}{\lower.7ex\hbox{$\;\stackrel{\textstyle<}{\sim}\;$}}

\def\i{\mathrm{i}}

\RequirePackage[normalem]{ulem}

\usepackage{orcidlink}

\usepackage{orcidlink}

\begin{document}

\title{Coupled Dark Energy and Dark Matter for DESI: \\
An Effective Guide to the Phantom Divide}

\author{Stefan Antusch\,\orcidlink{0000-0001-6120-9279}}
\email{stefan.antusch@unibas.ch}
\affiliation{Department of Physics, University of Basel, Klingelbergstrasse 82, CH-4056 Basel,
Switzerland}

\author{Stephen F. King\,\orcidlink{0000-0002-4351-7507}}
\email{king@soton.ac.uk}
\affiliation{School of Physics and Astronomy, University of Southampton,
SO17 1BJ Southampton, U.K.}

\author{Xin Wang\,\orcidlink{0000-0003-4292-460X}}
\email{xin.wang@unipd.it}
\affiliation{Dipartimento di Fisica e Astronomia ``Galileo Galilei'', Universit\`a degli Studi di Padova,
Via Francesco Marzolo 8, 35131 Padova, Italy}
\affiliation{INFN, Sezione di Padova, Via Francesco Marzolo 8, 35131 Padova, Italy}

\begin{abstract}
Motivated by the recent Dark Energy Spectroscopic Instrument (DESI) DR2 preference for dynamical dark energy, we study interacting dark energy models in which a canonical quintessence field couples to cold dark matter through a field-dependent mass $m(\phi)$. In such scenarios, the {\it effective} equation of state inferred under the assumption of non-interacting dark sectors, $w_{\rm eff}(z)$, can differ from the intrinsic scalar-field equation of state $w_\phi(z)$, making an {\it apparent} phantom crossing $w_{\rm eff}<-1$ possible without introducing a phantom scalar. We show that a viable realization of this mechanism requires the scalar field to originate from a frozen phase deep in the radiation era, in order for the effective coupling to remain sufficiently suppressed before recombination to evade cosmic microwave background constraints, and for the late-time evolution to become strong enough to reproduce the {\it apparent} behavior of $w_{\rm eff}(z)$ preferred by DESI. We identify the general conditions that allow these requirements to be satisfied simultaneously, and present an illustrative phenomenological realization in which $w_{\rm eff}(z)$ evolves from $w_{\rm eff}\approx -1.2$ at $z \approx 1.0$ to $w_{\rm eff}\approx -0.9$ at $z\approx 0.4$. These conditions and requirements serve as a guide for designing future models of this kind which can safely navigate the phantom divide at $w=-1$ in an {\it effective} way without phantom fields.

\end{abstract}

\date{\today}

\maketitle

\section{Introduction}
\label{sec:introduction}

The $\Lambda$CDM paradigm~\cite{Efstathiou:1990xe,Frieman:2008sn,Weinberg:2013agg}, featuring a cosmological constant ($\Lambda$) and cold dark matter (CDM) has provided a robust foundation for understanding the Universe and achieved impressive agreement with a broad set of cosmological observations~\cite{SupernovaSearchTeam:1998fmf,SupernovaCosmologyProject:1998vns,2dFGRSTeam:2002tzq,2dFGRS:2005yhx,Planck:2018vyg,eBOSS:2021pff,DES:2017myr,eBOSS:2020yzd,Heymans:2020gsg,Brout:2022vxf,DES:2024jxu}. Nevertheless, as the precision of cosmological measurements continues to improve, persistent tensions and potential deviations from 
$\Lambda$CDM, including the possible nature of 
dark energy (DE), 
have attracted growing attention~\cite{Abdalla:2022yfr,Perivolaropoulos:2021jda}. 

The recent baryon acoustic oscillation (BAO) measurements from the Dark Energy Spectroscopic Instrument (DESI) provide some of the most precise late-time determinations of the distance-redshift relation to date~\cite{ DESI:2024mwx,DESI:2025fii, DESI:2025zgx,DESI:2025qqy,  Elbers:2025vlz}. The DESI BAO data are well described by a flat $\Lambda$CDM model, yet the parameters preferred by BAO exhibit a mild  tension ($\sim 2.3\sigma$) with those inferred from the cosmic microwave background (CMB)~\cite{DESI:2025fii,DESI:2025zgx}. This tension is alleviated by DE with a time-evolving equation of state parametrized by $w_0$ and $w_a$ as $w(a) = w_0 + w_a(1-a)$ with $a$ being the scale factor. A solution in the quadrant with $w_0 > -1$ and $w_a < 0$ is preferred over $\Lambda$CDM at $3.1\sigma$ for the combination of DESI BAO
and CMB data~\cite{DESI:2025fii, DESI:2025zgx}.\footnote{Note, however, that $w_0>-1$ is generally correlated with a lower inferred $H_0$ within the $w_0 w_a$ parametrization, therefore the $H_0$ tension is not typically alleviated in this region of parameter space~\cite{Lee:2022cyh, Colgain:2025nzf}.} This motivates an equation of state for DE that crosses the so called ``phantom divide'' at the point where $w=-1$~\cite{DESI:2025fii}. This result provides concrete observational motivation for earlier dynamical DE scenarios~\cite{Wetterich:1994bg,Ratra:1987rm,Sahni:1999gb,Peebles:2002gy,Guo:2004vg,Amendola:1999er,Farrar:2003uw,Khoury:2003rn, Das:2005yj,Antusch:2008hj,Cai:2021wgv,Copeland:2006wr,Guo:2007zk,Cai:2009ht,Bull:2015stt,Zlatev:1998tr,Caldwell:2005tm}, including those with so called ``phantom'' scalar fields with negative or ``wrong sign'' kinetic terms~\cite{Caldwell:1999ew,Cline:2003gs,Feng:2004ad,Ludwick:2017tox,Cai:2025mas}. Following the DESI, there has been much renewed interest in dynamical DE scenarios, see, e.g.\  Refs.~\cite{Giare:2024gpk,Li:2024qso,Sabogal:2025mkp, Wolf:2025jed,Li:2025owk,  Dinda:2025iaq,deSouza:2025rhv,Akrami:2025zlb,Bayat:2025xfr,Chen:2025ywv,Li:2025ula,Ozulker:2025ehg, Silva:2025twg, Nojiri:2025uew, Thanankullaphong:2026anl,Gialamas:2025pwv,Chakraborty:2025syu,Bedroya:2025fwh, Wang:2025znm,Samanta:2025oqz,Nojiri:2025low,SanchezLopez:2025uzw,LaPenna:2026avs,Li:2026xaz,Odintsov:2026doe,Odintsov:2026doe,Nojiri:2026uvn}.\footnote{It should be noted that, while DESI DR2 has been widely interpreted as strengthening the case for dynamical DE, several subsequent reanalyses have argued that the statistical significance of this preference is sensitive to dataset combination, supernova calibration, and the choice of model-comparison criterion, with some Bayesian analyses finding that the preference can be substantially weakened or even disappear~\cite{RoyChoudhury:2024wri,RoyChoudhury:2025dhe,Ong:2026tta,Wang:2025bkk,RoyChoudhury:2025iis,Cheng:2025yue}. Related studies have also emphasized that the evidence for a dynamical DE signal from DESI data alone remains limited~\cite{Colgain:2025nzf,Colgain:2025fct}. }

Here we discuss a class of interacting dynamical DE models, where the DE component is a scalar (quintessence) field that couples to fermionic dark matter (DM) through a Yukawa-type interaction~\cite{Amendola:1999er,Farrar:2003uw,Khoury:2003rn, Das:2005yj,Antusch:2008hj,Cai:2021wgv,Chakraborty:2025syu}. Such a coupling induces an energy exchange within the dark sector, implying that the {\it effective} equation-of-state parameter, $w_{\rm eff}$, inferred by assuming non-interacting components, can differ from the {\it intrinsic} scalar-field equation of state, $w_\phi$. As a result, the {\it effective} DE phenomenology can naturally mimic an {\it apparent} phantom crossing, even though the scalar field itself has a normal ``correct sign'' positive kinetic term, unlike the phantom fields considered earlier.

In the present paper we emphasize that, within the above approach of non-phantom scalar field coupled to fermionic DM,  
achieving a significant evolution of $w_{\rm eff}$ from the ``effective phantom region'' to $w_{\rm eff}>-1$ today ($a=1$) typically requires the scalar field to evolve away from its minimum. We point out that this requirement demands a carefully controlled set of initial conditions. In particular, we show that the scalar dynamics must start from a frozen phase, characterized by an initial velocity with respect to e-fold time close to zero. By analyzing the structure of the field equation, we find that placing this frozen phase deep in the radiation-dominated (RD) era leads to a much more robust evolution, namely, the subsequent trajectory becomes insensitive to small perturbations in the initial velocity and avoids an unphysical kination regime under backward integration. At the same time, explaining the late-time behavior motivated by DESI without significantly violating CMB constraints imposes nontrivial restrictions on the forms of the scalar potential and the coupling function. 

We elucidate general conditions  that models of non-phantom scalar field coupled to fermionic DM should satisfy, and introduce 
local linear expansions of the scalar potential and the coupling function both around the era of recombination and around the present epoch. Guided by these considerations, we then construct a minimal phenomenological realization which satisfies conservative CMB-safe conditions and reproduces the binned phenomenological reconstruction of $w(z)$ in the DESI DR2 extended DE analysis~\cite{DESI:2025fii,DESI:2025zgx}. This minimal phenomenological realization can serve as a guide for building future particle physics motivated models which can safely navigate the phantom divide at $w=-1$ in an {\it effective} way without phantom fields.

The layout of this paper is as follows. In Sec.~\ref{sec:model}, we establish the general framework of the non-phantom quintessence scalar DE coupled to a fermionic DM field via a Yukawa interaction, leading to an {\it effective} equation of state capable of crossing the phantom divide. We also describe the qualitative scenario which we assume throughout the remainder of the paper.
In Sec.~\ref{sec:evolution}, we discuss the evolution of the {\it effective} DE and derive the general conditions that should be satisfied in order to yield the desired initial conditions corresponding to the frozen phase. We present a minimal realization in Sec.~\ref{sec:realization} and show that all these requirements are satisfied, within a simple phenomenological set-up. The main results are summarized in Sec.~\ref{sec:conclusion}, while Sec.~\ref{conclusion} concludes the paper.

\section{The Framework}
\label{sec:model}
The starting point is a (non-phantom) quintessence scalar field $\phi$, coupled to a fermionic Dirac DM field $\psi$ via a Yukawa interaction\footnote{This is analogous to the Higgs Yukawa coupling to the electron field in the Standard Model, at low energies, after spontaneous symmetry breaking, but allowing for a more general form of coupling.}
\begin{align}
    -{\cal L}_{\rm Y} = \mu f(\phi)\overline{\psi}\psi \; ,
    \label{eq:yukawa}
\end{align}
where $\mu$ is a positive mass scale, and $f(\phi)$ is a general dimensionless function, which we assume to be positive definite, to avoid vanishing DM mass that would be at odds with the cold DM paradigm that we assume here.

In the case of homogeneous, adiabatic, and non-relativistic cold DM, the fermion field $\psi$ gives rise to a condensate with a number density $n(a) = n_0/a^3 \simeq \langle \overline{\psi}\psi \rangle$.\footnote{Strictly speaking, the particle number density of $\psi$ should be $n \equiv \langle \psi^\dagger \psi \rangle $, whereas the Yukawa term in Eq.~\eqref{eq:yukawa} gives rise to a Lorentz scalar $n_s \equiv \langle \overline{\psi}\psi \rangle$. For a one-particle state with four-momentum $p^\mu = (E,\textbf{p})$, one has $n_s = n (m/E)$ with $m$ being the mass of $\psi$. Hence $n(a) \simeq \langle \overline{\psi}\psi \rangle$ is valid only for non-relativistic CDM with $E \simeq m$.} Then its energy density can be expressed as
\begin{align}
    \rho_{\rm DM} = m(\phi) \frac{n_0}{a^3} = \frac{\rho_{\mathrm{DM}}^{(0)}}{a^3}\frac{m(\phi)}{m_0} =\frac{\rho_{\mathrm{DM}}^{(0)}} {a^3}\frac{f(\phi)}{f_0} \; ,
    \label{DM}
\end{align}
where $m(\phi)$ denotes the $\phi$-dependent DM mass and the subscript/superscript ``$0$'' labels the values measured today. For simplicity, we assume that baryons and radiation are minimally coupled and separately conserved. In this case, due to the Yukawa interaction term, the physical action relevant for the scalar field $\phi$ in the Einstein frame should be revised as
\begin{align}
S = \int d^4x\sqrt{-g}\left[
\frac{M_{\rm Pl}^2}{2}R
-\frac12(\nabla\phi)^2 - V(\phi)
 - m(\phi)\,\bar\psi\psi
\right] \; ,
\end{align}
where $g\equiv \det(g_{\mu\nu})$ is the determinant of the spacetime metric, $M_{\rm Pl}$ denotes the reduced Planck mass, and $R$ is the Ricci scalar. In a spatially flat Friedmann-Robertson-Walker (FRW) background with a homogeneous field $\phi=\phi(t)$, varying the above action with respect to $\phi$ yields~\cite{Farrar:2003uw}
\begin{align}
    \ddot{\phi} + 3H\dot{\phi} + V_{\phi} 
    = -\frac{\partial m}{\partial \phi} \langle \bar{\psi}\psi \rangle 
    = -\frac{\partial}{\partial \phi} [ m(\phi) n(a) ] \; ,
    \label{eq:friedman-eq1}
\end{align}
which, from Eq.~\eqref{DM}, is equivalent to
\begin{align}
    \ddot{\phi} + 3H\dot{\phi}
    + \frac{\partial}{\partial\phi}
    \left[ V(\phi) + \frac{\rho_{\mathrm{DM}}^{(0)}} {a^3}\frac{f(\phi)}{f_0} \right]
    = 0 \; .
    \label{eq:friedman-eq2}
\end{align}
The term in square brackets defines an effective potential
\begin{align}
    V_{\rm eff} =  V(\phi) + \frac{\rho_{\mathrm{DM}}^{(0)}} {a^3}\frac{f(\phi)}{f_0} \; ,
    \label{eq:potential}
\end{align}
which consists of a sum of two terms: the ``bare'' scalar potential $V(\phi)$ which is assumed to dominate at the present time $a=1$,
and a contribution from DM which may dominate at early times when $a\ll 1$.

\begin{figure}[t!]
        \centering
        \includegraphics[width=1\linewidth]{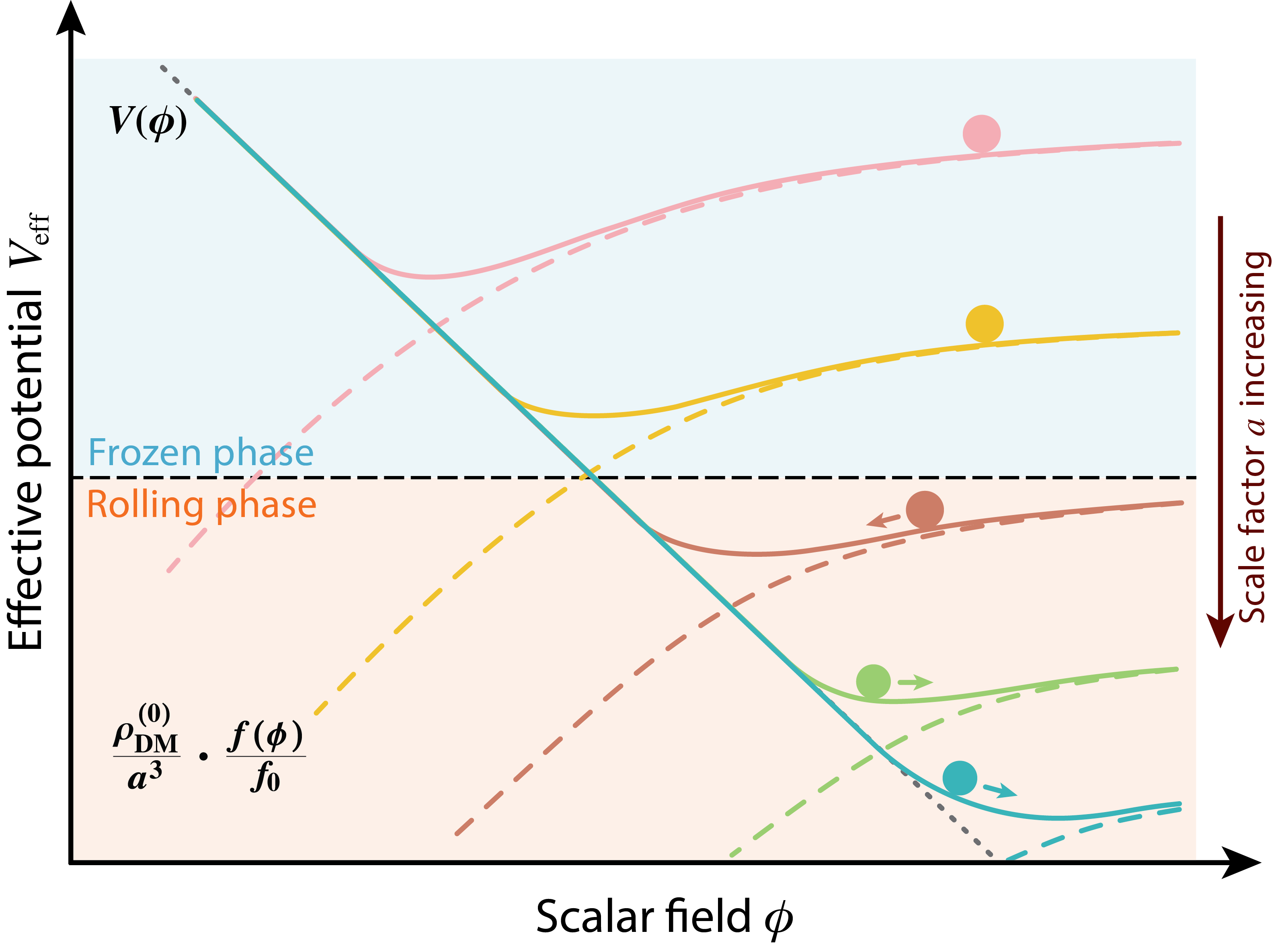}
        \caption{
        Schematic illustration of the scenario considered in this paper. In the shown example, both $V(\phi)$
        (short gray dotted line linearly decreasing)
        and $f(\phi)$ (long dashed colored lines and scaled by $1/a^{3}$) vary monotonically in the field range of interest.
        The effective potential (colored lines for different values of $a$) is the sum of these terms, along which the field $\phi$ rolls (filled circles), with arrows indicating the direction of motion of $\phi$. The field initially remains effectively frozen near its starting position at small $a$ and only begins to roll during the later-time evolution.
}
        \label{fig:schematic}
\end{figure}

Within this framework, the Hubble parameter can be determined by
\begin{align}
    3H^{2} M_{\mathrm{Pl}}^{2}
    = \rho_{\phi}
    + \frac{\rho_{\mathrm{DM}}^{(0)}}{a^{3}} \frac{f(\phi)}{f_{0}}
    + \frac{\rho_{\mathrm{B}}^{(0)}}{a^{3}} + \frac{\rho_{\rm r}^{(0)}}{a^4} \; ,
    \label{eq:hubble1}
\end{align}
where $\rho_\phi$, $\rho_{\rm B}$ and $\rho_{\rm r}$ denote the energy densities of the scalar field, baryons and radiation, respectively. It should be mentioned that while $\rho_{\rm r}$ is negligible at present, it becomes important when the initial conditions are imposed in the RD epoch. The effect of the interaction between DM and DE may be interpreted as an {\it effective} DE term, $\rho_{\rm DE}^{\rm eff}$, together with a standard DM evolution, as follows,
\begin{align}
    3H^{2} M_{\mathrm{Pl}}^{2}
    = \rho_{\rm DE}^{\rm eff}
    + \frac{\rho_{\mathrm{DM}}^{(0)}}{a^{3}} 
    + \frac{\rho_{\mathrm{B}}^{(0)}}{a^{3}} + \frac{\rho_{\rm r}^{(0)}}{a^4} \; .
    \label{eq:hubble2}
\end{align}
Comparing Eq.~\eqref{eq:hubble1} with Eq.~\eqref{eq:hubble2}, one can establish
\begin{align}
    \rho^{\rm eff}_{\rm DE} = \rho_\phi + \frac{\rho_{\rm DM}^{(0)}}{a^3} \left(\frac{f(\phi)}{f_0} - 1 \right) \; ,
    \label{eq:rhoDE_eff}
\end{align}
which satisfies an {\it effective} DE evolution equation 
involving an {\it effective} DM equation of state
$w_{\mathrm{eff}}$, 
\begin{align}
    \frac{\mathrm d}{\mathrm{d}t}
    \rho_{\mathrm{DE}}^{\mathrm{eff}} 
    + 3H \left( 1 + w_{\mathrm{eff}} \right)
      \rho_{\mathrm{DE}}^{\mathrm{eff}}
    = 0 \; .
\end{align}

From Eq.~\eqref{eq:rhoDE_eff} it is evident that $\rho_{\rm DE}^{\rm eff}$ includes an additional contribution arising from the DM sector. However, 
since the DM field is pressureless, the pressure term originates solely from the $\phi$ field, implying $\rho_{\rm DE}^{\rm eff} w_{\rm eff} = p_\phi = \rho_\phi w_\phi$. We thus have
\begin{align}
    w_{\mathrm{eff}} &= \frac{w_{\phi}}{1 - x} \; ,
    \label{eq:weff}
\end{align}
with
\begin{align}
    x & \equiv -\,\frac{\rho_{\mathrm{DM}}^{0}}{a^{3}\rho_{\phi}}
        \left( \frac{f(\phi)}{f_{0}} - 1 \right)\; , \quad
    w_{\phi} \equiv 
    \frac{\dot{\phi}^{2}/2 - V(\phi)}
         {\dot{\phi}^{2}/2 + V(\phi)} \; .
         \label{eq:weff-notation}
\end{align}

Before turning to the more quantitative discussion of the evolution in the next section, it is useful to summarize the qualitative picture implied by this framework. Near the present epoch, where the bare potential $V(\phi)$ dominates the effective potential in Eq.~\eqref{eq:potential}, we take the field to roll along the direction of decreasing $V(\phi)$, so that $V'(\phi_0)<0$ and $\phi$ evolves toward larger values. If the inferred $w_{\rm eff}$ is to exhibit an {\it apparent} phantom-crossing behavior at low redshifts, the DM mass at earlier times must be smaller than its present value, namely $f(\phi)<f_0$, so that $x$ in Eq.~\eqref{eq:weff-notation} is positive. In realizations where both $V(\phi)$ and $f(\phi)$ vary monotonically, this typically corresponds to a local slope of $f(\phi)$ opposite in sign to $V'(\phi)$ over the relevant field range. Due to the dependence on $1/a^3$, the interaction term proportional to $f(\phi)/a^3$ generally becomes increasingly important toward the past, and the effective potential can develop a time-dependent minimum $\phi_{\min}(a)$. If the scalar field does not adiabatically track this evolving minimum, its trajectory can pass through $\phi_{\min}(a)$ as it shifts with time, which generically leads to a change in the sign of $\dot\phi$. 
This qualitative picture is illustrated schematically in Fig.~\ref{fig:schematic}, and will motivate the more quantitative analysis in the next section.

\section{The Evolution of Dark Energy}
\label{sec:evolution}
As discussed qualitatively in the previous section, our aim is to realize a robust DE evolution that starts from a frozen phase, where $\phi$ rolls very slowly on a Hubble timescale, i.e.\ the scalar motion is strongly damped by the Hubble friction. We further require this set-up to generate $w_{\rm eff}$ exhibiting a phantom-crossing behavior around $z\sim \mathcal{O}(1)$, while satisfying the early-Universe constraints by keeping both the DE fraction and the deviation from the standard DM evolution sufficiently suppressed near recombination. 

\subsection{Linear Parameterization} 

In this section we extract some quantitative and general (model-independent) guidance from these requirements. To this end, the potential $V(\phi)$ and the interaction function $f(\phi)$ may be expanded around any arbitrary field value $\phi_*$, to linear order in the local linear expansions as
\begin{align}
V(\phi)  \simeq V_*+ \alpha_*\,\Delta\phi \; , \quad
f(\phi) \simeq f_*(1+\beta_*\Delta\phi) \; ,
\label{eq:toy_linear_alpha_neg}
\end{align}
where  $\alpha_* \equiv ({\rm d}V/{\rm d}{\phi})|_{\phi_*}$, $\beta_* \equiv ({\rm d}\,{\rm ln}\,f/{\rm d}{\phi})|_{\phi_*}$, and $\Delta \phi \equiv \phi - \phi_*$, while $V_*$ and $f_*$ are the respective values of $V(\phi)$ and $f(\phi)$ computed at $\phi_*$. Without loss of generality, we shall adopt the convention that $\alpha_* < 0$.

So far this is just a mathematical formalism, a first order approximation to a Taylor expansion around some arbitrary $\phi_*$.  
In the following, we will perform such linear expansions of $f(\phi)$ and $V(\phi)$ at different epochs of interest, and accordingly $\phi_*$ will be chosen to be a reference field value corresponding to the different epochs (times) of interest, and in each case,
$\Delta\phi$ will be the deviation as the field evolves in time. The accuracy of the linear approximation will need to be examined for each case. The key 
feature of this bottom-up approach to the unknown potential and coupling is that the values of
 $V_*$ and $f_*$, as well as $\alpha_*$ and
$\beta_*$, are taken to be free parameters, with different values at the different epochs of interest: late time, early time and CMB or recombination era (as defined and discussed below).

It is intuitive to use the e-fold time $N \equiv \ln a$ and define $\phi^\prime \equiv {\rm d}\phi/{\rm d}\ln a = \dot \phi / H$. Then Eq.~\eqref{eq:friedman-eq2} can be rewritten as
\begin{align}
    \phi^{\prime\prime} + (3 - \epsilon_H) \phi^\prime + \frac{1}{H^2}\left( V_\phi + \frac{\rho_{\mathrm{DM}}^{(0)}} {a^3}\frac{f_\phi}{f_0}\right) = 0 \; ,
    \label{eq:friedman-eq3}
    \end{align}
where $\epsilon_H \equiv -{\rm d}\ln H/{\rm d}\ln a$. Within the local expansion in Eq.~\eqref{eq:toy_linear_alpha_neg}, this reduces to
\begin{align}
    \phi^{\prime\prime} + (3 - \epsilon_H) \phi^\prime + \frac{\alpha_*}{H^2} + \frac{\rho_{\mathrm{DM}}^{(0)}} {a^3 H^2}\frac{f_*}{f_0}\beta_* \simeq 0 \; .
    \label{eq:KGequ-approx}
\end{align}
In the following, we use the linearized
Eq.~\eqref{eq:KGequ-approx}
to discuss late-time fitting, early-time freezing, and the qualitative CMB requirements in a unified way.

\subsection{Late-time fitting}  

We first consider the present epoch, where, according to our scenario, the field is slowly rolling and the evolution is overdamped, so one may neglect $\phi''$ in Eqs.~\eqref{eq:friedman-eq3} and \eqref{eq:KGequ-approx} to the first approximation. In this case, we will expand around $\phi_*=\phi_0$, the present day field value, in order to find $\Delta\phi$ in the recent past at {\it late times} in the history of the Universe. 

Expanding around $\phi_0$, we obtain from Eq.~\eqref{eq:KGequ-approx},
\begin{align}
3H^2\phi'&\simeq-\left(\alpha_0+\beta_0\,\rho^{(0)}_{\rm DM}a^{-3} \right)\;.
\label{eq:slowroll}
\end{align}
The kinetic energy is therefore
\begin{align}
K=\frac{\dot\phi^2}{2}=\frac{H^2\phi'^2}{2} &\simeq \frac{\left(\alpha_0+\beta_0\rho^{(0)}_{\rm DM}a^{-3}\right)^2}{18H^2}\; , 
\end{align}
which yields
\begin{align}
1+w_\phi &\simeq \frac{\left(\alpha_0+\beta_0\rho^{(0)}_{\rm DM}a^{-3}\right)^2}{9H^2V(\phi)}\;.
\end{align}
On the other hand, the quantity $x$ that controls the difference between $w_{\rm eff}$ and $w_\phi$ can be estimated as
\begin{align}
x \simeq -\frac{\rho^{(0)}_{\rm DM}}{a^3\rho_\phi}\beta_0\Delta\phi \; ,
\end{align}
where $|\beta_0\Delta\phi| \ll 1$ has been taken into account. We assume $V(\phi)$ dominates the total effective potential today, hence the field is rolling towards larger values, which indicates
$\Delta \phi(a<1)<0$. Then in order to obtain 
$w_{\rm eff}<-1$, we require $x>0$, which in turn implies $\beta_0>0$. In this sense, the slopes of $V(\phi)$ and $f(\phi)$ are necessarily opposite. Moreover, $\alpha_0$ mainly fixes $1 + w_\phi$ at the lowest redshifts, while $\beta_0$ controls how deeply $w_{\rm eff}$ enters the {\it apparent} phantom region.

We approximate the late-time expansion by
\begin{align}
H^2(a)&\simeq H_0^2\left(\Omega_{\rm DE0}+\Omega_{\rm m0}a^{-3}\right)\; ,
\end{align}
where $\Omega_{\rm DE0}$ and $\Omega_{\rm m0}$ denote the energy fraction of DE and matter today, satisfying $\Omega_{\rm DE0} + \Omega_{\rm m0} \simeq 1$. Keeping this in mind, Eq.~\eqref{eq:slowroll} can be integrated as
\begin{align}
\Delta\phi(a)\simeq
&-\frac{\alpha_0}{9H_0^2\Omega_{\rm DE0}}
\ln\!\left(
\Omega_{\rm DE0}a^3+\Omega_{\rm m0}
\right)\nonumber\\
&+\frac{\beta_0\rho^{(0)}_{\rm DM}}{9H_0^2\Omega_{\rm m0}}
\ln\!\left(
\Omega_{\rm DE0}+\Omega_{\rm m0}a^{-3}
\right)\;,
\label{eq:displacement}
\end{align}
which indicates that, both $\alpha_0$ and $\beta_0$ should in general contribute to the field displacement.

We implement an order-of-magnitude estimation of $\alpha_0$ and $\beta_0$ by leveraging the binned phenomenological reconstruction of $w(z)$ using DESI+CMB+Union3 in the DESI DR2 extended DE analysis~\cite{DESI:2025fii,DESI:2025zgx}. For convenience, we define the following dimensionless quantities
\begin{align}
\widehat{\alpha}_0 \equiv \frac{\alpha_0}{H_0^2M_{\rm Pl}}\;,\quad
\widehat{\beta}_0 \equiv \beta_0 M_{\rm Pl}\;, \quad
\Delta(a) \equiv \frac{\Delta\phi(a)}{M_{\rm Pl}}\;.
\end{align}
In the $V(\phi)$-dominance regime, $V_0 \simeq 3H^2_0 M^2_{\rm Pl}\Omega_{\rm DE0}$. We further assume the bare slope dominates the DM-induced slope, $|\alpha_0| \gg \rho_{\rm DM}^{(0)} |\beta_0| / a^3$, which is valid at least for the lowest redshift bin. Then Eq.~\eqref{eq:displacement} reduces to
\begin{align}
\Delta(a)\simeq
&-\frac{\widehat{\alpha}_0}{9\Omega_{\rm DE0}}
\ln\!\left(
\Omega_{\rm DE0}a^3+\Omega_{\rm m0}\right)
 \;,
 \label{eq:dimless-displacement}
\end{align}
Correspondingly, one has
\begin{align}
\label{eq:lin_wphi}
1+w_\phi(a) & \simeq \frac{\widehat{\alpha}^2_0}
{27\Omega_{\rm DE0}(\Omega_{\rm DE0}+\Omega_{\rm m0}a^{-3})}\;, \\
\label{eq: lin_x}
x(a) & \simeq -\frac{\Omega_{\text{DM}0}\widehat{\beta}_0\Delta(a)}{\Omega_{\rm DE0}a^3}\;,
\end{align}
where $\Omega_{\text{DM}0}$ denotes the present DM fraction. 

For illustration, we use the first two reconstructed DESI bins for the  calibration of $\alpha_0$ and $\beta_0$, namely,
\begin{align}
w_{\rm eff}(a_1
\simeq 0.75)\simeq -0.9\;,\quad
w_{\rm eff}(a_2 \simeq 0.5)\simeq -1.35\;, \nonumber
\end{align}
and $(\Omega_{\rm m0},\Omega_{\rm DE0},\Omega_{{\rm DM}0})=(0.3,0.7,0.25)$. As we can see, $a_1$ is  close to unity, hence we assume $w_{\rm eff} \simeq w_\phi$. Then $\widehat{\alpha}_0$ is uniquely determined from the first bin using Eq.~\eqref{eq:lin_wphi}, and $\widehat{\beta}_0$ can be subsequently obtained from the second bin using Eqs.~\eqref{eq:dimless-displacement}--\eqref{eq: lin_x}, together with the relation $w_{\rm eff} = w_\phi/(1-x)$. As a result, we get
\begin{align}
\widehat{\alpha}_0\simeq -1.6\;,\qquad \widehat{\beta}_0\simeq 0.4\;.
\end{align}

It is worth mentioning that the above analysis can be regarded only as a qualitative estimation. In fact, the assumption that the DM-induced slope is parametrically subdominant no longer holds at $z\simeq 1$.
Therefore, the $V_\phi$-dominance assumption is at best marginally consistent for this numerical example, and a more accurate treatment should keep the DM-induced term in the slow-roll equation for $\phi$. As we shall see later, we can indeed achieve a good fit for the first two redshift bins by tuning the parameters numerically.

\subsection {Early-time freezing} 

For a more complete cosmological history of the Universe, we also need to consider {\it early times} in the radiation dominated (RD) era.
We therefore need to identify a physical solution that satisfies Eq.~\eqref{eq:friedman-eq3} and connects some initial conditions to the late-time behavior discussed above. It is worth mentioning that for constant $\epsilon_H$, Eq.~\eqref{eq:friedman-eq3} contains a homogeneous mode proportional to $a^{-(3-\epsilon_H)}$, which is strongly damped by Hubble friction in forward time evolution, but grows rapidly under backward integration. Hence, reconstructing the past evolution of $w_{\rm eff}$ by integrating backward from the field velocity and energy density at the present time is in general unstable, since even a tiny numerical mismatch is amplified toward the past. A more robust strategy is therefore to impose the initial condition in the early Universe and evolve the system forward in time. The question is then under what conditions an approximately frozen initial velocity, as assumed on naturalness grounds in our scenario, can remain compatible with the general solution.

In the RD era, $\epsilon_H \simeq 2$ and $H^2 \simeq H_0^2\Omega_{\rm r0}a^{-4}$, with $\Omega_{\rm r0}$ being the radiation energy fraction today. We therefore set $\phi_*=\phi_{\rm ini}$ and $\Delta\phi = \phi - \phi_{\rm ini}$, where the subscripts ``ini'' refer to the initial field value. The linear expansions in Eq.~\eqref{eq:toy_linear_alpha_neg} then become
\begin{align}
V(\phi) \simeq V_{\rm ini} + \alpha_{\rm ini}\Delta\phi\;, \quad
f(\phi)\simeq f_{\rm ini}(1+\beta_{\rm ini}\Delta\phi)\;.
\end{align}
Then Eq.~\eqref{eq:KGequ-approx} yields,
\begin{align}
\phi''+\phi' +A_{\rm ini}\,a +B_{\rm ini}\,a^4 \simeq 0\;,
\label{eq:RDequation}
\end{align}
where
\begin{align}
A_{\rm ini} \equiv \frac{3\Omega_{{\rm DM}0}}{\Omega_{\rm r0}}\frac{f_{\rm ini}}{f_0}\widehat{\beta}_{\rm ini} M_{\rm Pl}\;,\quad
B_{\rm ini}\equiv
\frac{\widehat{\alpha}_{\rm ini}M_{\rm Pl}}{\Omega_{\rm r0}}\;,
\label{eq:RDcoeff}
\end{align}
with $\widehat{\alpha}_{\rm ini}\equiv \alpha_{\rm ini}/(H^2_0 M_{\rm Pl})$ and $\widehat{\beta}_{\rm ini}\equiv \beta_{\rm ini}M_{\rm Pl}$. The exact solution to Eq.~\eqref{eq:RDequation} is
\begin{align}
\phi'_{\rm RD}(a) = C a^{-1} -\frac{A_{\rm ini}}{2}a
-\frac{B_{\rm ini}}{5}a^4\;,
\label{eq:RDgeneral}
\end{align}
where $C$ is an integration constant. A detailed derivation of Eq.~\eqref{eq:RDgeneral} can be found in appendix~\ref{app:derivation}. The first term is the homogeneous mode, while the remaining terms arise from the derivative of the effective potential. For an initial condition specified at some finite $a_{\rm ini}$, the coefficient $C$ is fixed by
\begin{align}
C = a_{\rm ini}\left[\phi'(a_{\rm ini})+\frac{A_{\rm ini}}{2}a_{\rm ini} +\frac{B_{\rm ini}}{5}a^4_{\rm ini}\right]\;.
\end{align}

It is useful to identify the last two terms on the right-hand side of Eq.~\eqref{eq:RDgeneral} as a special branch, namely,
\begin{align}
\phi'_{\rm att,RD}(a)= -\frac{A_{\rm ini}}{2}a
-\frac{B_{\rm ini}}{5}a^4\;,
\end{align}
which tends to zero as $a \to 0$. Accordingly, Eq.~\eqref{eq:RDgeneral} can be rewritten as
\begin{align}
\delta\phi'_{\rm RD}(a)=\delta\phi'_{\rm RD}(a_{\rm ini})\frac{a_{\rm ini}}{a}\;,
\end{align}
where
\begin{align}
\delta\phi'_{\rm RD}(a)\equiv \phi'(a)-\phi'_{\rm att,RD}(a)\;.
\end{align}
This explicitly shows that any initial mismatch with respect to the $\phi'_{\rm att,RD}(a)$ branch decays in forward evolution, indicating that $\phi'_{\rm att,RD}(a)$ is actually an attractor-like solution to the field equation. In this sense, the deep-RD solution naturally approaches a frozen configuration. To be more specific, we can initialize the field at a sufficiently early time, $a_{\rm ini}\ll 1$, with an approximately frozen velocity, $\phi'(a_{\rm ini})\simeq 0$. The reason is twofold. On the one hand, in the deep RD era the special branch $\phi'_{\rm att,RD}$ itself is already arbitrarily close to zero. On the other hand, if one chooses $\phi'(a_{\rm ini})\simeq 0$, then the initial mismatch is
\begin{align}
\delta\phi'_{\rm RD}(a_{\rm ini}) \simeq -\phi'_{\rm att,RD}(a_{\rm ini})\;,
\end{align}
which is small provided that $a_{\rm ini}$ is sufficiently small, and it continues to decrease during forward evolution. Therefore, a nearly frozen initial condition at sufficiently small $a_{\rm ini}$ is effectively equivalent to initializing the system near the RD attractor.

Moreover, in the $a\to 0$ limit $\phi'_{\rm att,RD}(a)$ becomes insensitive to the detailed global forms of the potential and the interaction, although it still depends on the local coefficients $A_{\rm ini}$ and $B_{\rm ini}$. For small $a$, the $A_{\rm ini}a$ term dominates over the $B_{\rm ini}a^4$ term over a wide interval, so the early departure from the frozen regime is typically driven first by the interaction term, while the potential contribution remains subleading. A sufficient condition for an approximately frozen start is therefore
\begin{align}
\left|\frac{\phi'_{\rm att,RD}(a_{\rm ini})}{M_{\rm Pl}}\right|
\simeq
\left|
\frac{3\Omega_{{\rm DM}0}}{2\Omega_{\rm r0}}
\frac{f_{\rm ini}}{f_0}\widehat{\beta}_{\rm ini}
\right|a_{\rm ini}
\ll1\;,
\end{align}
where the subleading term proportional to $B_{\rm ini}a_{\rm ini}^4$ has been neglected. Given that $\Omega_{{\rm DM}0} \simeq 0.25$ and $\Omega_{\rm r0} \simeq 9 \times 10^{-5}$, and assuming that $f_{\rm ini}$ and $f_0$ are of the same order, this condition implies
\begin{align}
a_{\rm ini}\widehat{\beta}_{\rm ini}\ll 2.4\times 10^{-4}\;,
\end{align}
which makes it clear that the existence of a frozen initial velocity is controlled not by $\widehat{\beta}_{\rm ini}$ alone, but by the combination $a_{\rm ini}\widehat{\beta}_{\rm ini}$, provided that the initial time is chosen in the RD era.

For comparison, during matter domination (MD) one has $\epsilon_H \simeq 3/2$ and $H^2 \simeq H_0^2\Omega_{\rm m0}a^{-3}$. Under the same local linear expansion, the field equation becomes
\begin{align}
\phi''+\frac{3}{2}\phi' +A_{\rm ini}^\prime +B_{\rm ini}^\prime \,a^3 \simeq 0\;,
\label{eq:MDequation}
\end{align}
where
\begin{align}
A_{\rm ini}^\prime \equiv 
\frac{3\Omega_{{\rm DM}0}}{\Omega_{\rm m0}}\frac{f_{\rm ini}}{f_0}\widehat{\beta}_{\rm ini}M_{\rm Pl}\;,\quad
B'_{\rm ini}\equiv
\frac{\widehat{\alpha}_{\rm ini}M_{\rm Pl}}{\Omega_{\rm m0}}\;.
\end{align}
The exact general solution to Eq.~\eqref{eq:MDequation} is
\begin{align}
\phi'_{\rm MD}(a)
=
C'a^{-3/2}
-\frac{2}{3}A^\prime_{\rm ini}
-\frac{2}{9}B^\prime_{\rm ini}a^3\;,
\label{eq:MDgeneral}
\end{align}
where $C'$ is an integration constant. Hence the homogeneous contribution again decays in forward evolution. Defining
\begin{align}
\phi'_{\rm att,MD}(a)\equiv
-\frac{2}{3}A^\prime_{\rm ini}
-\frac{2}{9}B^\prime_{\rm ini}a^3\;,
\end{align}
one finds
\begin{align}
\delta\phi'_{\rm MD}(a) = \delta\phi'_{\rm MD}(a_{\rm ini})\left(\frac{a_{\rm ini}}{a}\right)^{3/2}\;,
\end{align}
with
\begin{align}
\delta\phi'_{\rm MD}(a)\equiv \phi'(a)-\phi'_{\rm att,MD}(a)\;.
\end{align}
Thus the MD dynamics is also stable in the sense that the homogeneous mode still decays. 
The essential difference from the RD case is that $ \phi'_{\rm att,MD}(a)$ is not asymptotically frozen. Its leading term is the nonzero constant $-2A^\prime_{\rm ini}/3$. Therefore, if one still insists on imposing a nearly frozen initial condition, $\phi'(a_{\rm ini})\simeq 0$, the initial condition is generally not close to the MD branch unless $A^\prime_{\rm ini}$ is already sufficiently small. In other words, although the homogeneous mode still decays in the MD era, a frozen initial phase no longer arises as a robust late-time continuation of the general solution. Instead, it becomes a tuned choice unless the interaction-induced term is already strongly suppressed.

In summary, a local linear expansion around the initial field value allows one to investigate the early-time scalar-field equation in a model-independent way. The general solution contains a homogeneous mode that diverges toward the past as $a\to 0$, indicating that a backward reconstruction from late times is generally unstable and can easily drive the field into a kinetic-dominated regime. This makes it more robust to impose the initial condition in the early Universe and evolve forward in time. The RD and MD attractors derived from the general solution have qualitatively different structures. In the MD era the attractor is usually controlled by a nonzero constant term, whereas in the RD era it approaches zero in the limit $a\to 0$. It is therefore the deep RD era that naturally provides the appropriate window for implementing a frozen initial condition in a robust way.

\subsection{CMB requirements} 

Successful models must satisfy cosmological constraints not only at the background level but also at the perturbation level. While a full Boltzmann analysis including perturbations is beyond the scope of this work, one can still extract several conservative necessary conditions from the CMB safety requirement.

First, since the Yukawa coupling makes the CDM mass depend on $\phi$, a non-vanishing $\dot{\phi}$ at early times opens an energy-transfer channel between the scalar and CDM. Since $\rho_{\rm DM} = m(\phi)\, n$, with $m(\phi)\equiv\mu f(\phi)$ the DM mass and $n(a)\propto a^{-3}$ the DM number density, the background continuity equations for DE and DM can be written as
\begin{align}
\dot{\rho}_{\phi}+3H(1+w_{\phi})\rho_{\phi} &= -\dot m(\phi)\, n \equiv +Q \;, \\
\dot{\rho}_{\rm DM}+3H\rho_{\rm DM} &= +\dot m(\phi)\, n \equiv -Q \; ,
\end{align}
where the energy transfer rate $Q$ is hence
\begin{align}
    Q = -\,\rho_{\rm DM} \frac{d\ln m}{d\phi}\dot{\phi} \; ,
\end{align} 
with the sign convention that $Q>0$ corresponds to energy transfer from DM to DE. The energy transfer in the dark sector can induce relative-density fluctuations between components, i.e.\ non-adiabatic isocurvature perturbations. These modes, if present during the deep radiation era, can affect the CMB acoustic pattern and the subsequent growth of structure. Consequently, CMB consistency requires the interaction rate to be sufficiently small around $z\sim 10^3$, thereby imposing an upper bound on the energy exchange during recombination. We introduce a dimensionless parameter $\epsilon(z)$ to quantify the interaction strength
\begin{align}
\epsilon\equiv\frac{|Q|}{H\rho_{\rm DM}}
 \simeq \left|\frac{{\rm d}\ln f}{{\rm d}\phi}\right|\,|\phi'| \; .
\end{align}

Following a similar strategy to the previous subsections, we now choose $\phi_*$ to be the field value at the recombination epoch $\phi_{\rm rec}$, and expand $f(\phi)$ as $f(\phi)\simeq f_{\rm rec}(1+\beta_{\rm rec}\Delta \phi)$, where $\beta_{\rm rec} \simeq ({\rm d}\ln f/{\rm d}\phi)_{\rm rec}$ and $\Delta\phi = \phi - \phi_{\rm rec}$. Physically, recombination corresponds to a finite redshift interval. However, the CMB is mainly sensitive to the relatively narrow time window around $z\sim 10^3$. For definiteness, we take the photon-decoupling redshift $z_*\simeq 1090$ as the representative redshift defining $\phi_{\rm rec}$. We also note that $\phi$ should vary extremely slowly during recombination, so the precise choice of this reference redshift has a negligible effect on the linear expansion. 

As discussed above, the DM-induced slope $f(\phi)$ dominates over the potential $V(\phi)$ in the field equation at early times. In the slow-roll regime, the field equation 
Eq.~\eqref{eq:friedman-eq2} is then
\begin{align}
3H\dot\phi \simeq -\frac{\rho_{\rm DM0}}{a^3}\frac{f_{\rm rec}}{f_0}\beta_{\rm rec}
\simeq -\frac{\rho_{\rm DM0}}{a^3}\beta_{\rm rec}\; ,
\end{align}
where in the last step we used $f_{\rm rec}\simeq f_0$. This approximation is physically well motivated, as a sizable deviation of $f_{\rm rec}$ from $f_0$ would imply an appreciable change in the DM mass since recombination, which is generally disfavored by CMB constraints on the matter density and expansion history. We then obtain
\begin{align}
\frac{\phi'}{M_{\rm Pl}}
= \frac{\dot\phi}{H M_{\rm Pl}}
\simeq -\,\frac{\Omega_{{\rm DM}0}\widehat{\beta}_{\rm rec}}{\Omega_{\rm m0}+\Omega_{\rm r0}/a} \; ,
\label{eq:phiprime-CMB}
\end{align}
where $\widehat{\beta}_{\rm rec}\equiv \beta_{\rm rec}M_{\rm Pl}$. At recombination, $z_{\rm rec}\simeq 1090$, hence $a_{\rm rec}=(1+z_{\rm rec})^{-1}\simeq 9.16\times 10^{-4}$. Substituting the fiducial present-day fractions $(\Omega_{\rm m0},\Omega_{{\rm DM}0},\Omega_{\rm r0})=(0.3,0.25,9\times 10^{-5})$, one finds
\begin{align}
\epsilon_{\rm rec}
\simeq \widehat{\beta}_{\rm rec}\left|\frac{\phi'(z_{\rm rec})}{M_{\rm Pl}}\right|
\simeq 0.625\,\widehat{\beta}_{\rm rec}^2 \; .
\end{align}
The CMB-safe requirement leads to a heuristic constraint $\epsilon_{\rm rec}\lesssim {\rm few}\times 10^{-3}$~\cite{Li:2024qso, Costa:2016tpb}, which implies $\widehat{\beta}_{\rm rec} \lesssim {\cal O}(0.1)$. This is much smaller than the late-time value of $\widehat{\beta}_0$ estimated above. 

Second, the cumulative variation of the CDM mass before recombination must remain perturbatively small, so that CDM still redshifts approximately as $a^{-3}$ over the pre-recombination era. In the local expansion around recombination, this requires
\begin{align}
|\Delta \ln m|_{\rm rec} \equiv \left|\ln\frac{m(\phi_{\rm rec})}{m(\phi)}\right|\simeq \left|\widehat{\beta}_{\rm rec}\frac{\Delta\phi_{\rm rec}(a)}{M_{\rm Pl}}\right|\ll 1 \; ,
\end{align}
where $\Delta\phi_{\rm rec}(a)\equiv \phi_{\rm rec}-\phi(a)$ with $a<a_{\rm rec}$, and we have used $|\beta_{\rm rec}\Delta \phi_{\rm rec}|\ll 1$. In particular, we examine the DM mass drift between matter–radiation equality $a_{\rm eq} \equiv \Omega_{\rm r0}/\Omega_{\rm m0}$ and recombination $a_{\rm rec}$. Using Eq.~\eqref{eq:phiprime-CMB}, this becomes
\begin{align}
\widehat{\beta}_{\rm rec}\frac{\Delta\phi_{\rm rec}(a_{\rm eq})}{M_{\rm Pl}} & \simeq \widehat{\beta}_{\rm rec}\int_{\ln a_{\rm eq}}^{\ln a_{\rm rec}}\frac{\phi'}{M_{\rm Pl}}\,{\rm d}\ln a \nonumber \\
& \simeq -\widehat{\beta}_{\rm rec}^2
\frac{\Omega_{{\rm DM}0}}{\Omega_{\rm m0}}
\ln\!\left(1 + \frac{a_{\rm rec}}{a_{\rm eq}}
\right) \; .
\end{align}
Again we use $(\Omega_{\rm m0},\Omega_{{\rm DM}0},\Omega_{\rm r0})=(0.3,0.25,9\times 10^{-5})$ for illustration, which leads to
\begin{align}
    \frac{\Omega_{{\rm DM}0}}{\Omega_{\rm m0}}
\ln\!\left(1 + \frac{a_{\rm rec}}{a_{\rm eq}}
\right) \simeq 1.17 \; . \nonumber
\end{align}
We require that the cumulative DM mass drift prior to recombination does not exceed approximately 1\%, i.e.\ $|\Delta \ln m|_{\rm rec} \lesssim 10^{-2}$. Then an upper bound $\widehat{\beta}_{\rm rec} \lesssim {\cal O}(0.1)$ can be derived.

Finally, the scalar field must provide a subdominant contribution to the total energy density at recombination. As a benchmark, we adopt the Planck 2015 upper bound on a constant early-DE fraction~\cite{Planck:2015bue}, according to which the energy fraction $\Omega_\phi$ of the scalar field at recombination should satisfy
\begin{align}
\Omega_\phi(z_{\rm rec}) < 0.0036 \quad (95\%~{\rm C.L.}) \; .
\end{align}
We note that as for the two constraints discussed above, also the value used here should be considered as a conservative necessary proxy only.
Since the potential energy is negligible at that epoch, $\Omega_\phi$ reduces to the kinetic estimate
\begin{align}
\Omega_\phi(z_{\rm rec})
\simeq
\frac{\dot\phi^2_{\rm rec}}{6H^2_{\rm rec}M^2_{\rm Pl}}
\simeq
\frac{1}{6}
\left(
\frac{\Omega_{{\rm DM}0}\widehat{\beta}_{\rm rec}}
{\Omega_{\rm m0}+\Omega_{\rm r0}/a_{\rm rec}}
\right)^2 \; ,
\end{align}
which indicates $\widehat{\beta}_{\rm rec} \lesssim 0.08$.

Overall, these three conditions impose  upper bounds on the linear expansion coefficient of $f(\phi)$ around recombination, requiring $\widehat{\beta}_{\rm rec}$ to be significantly smaller than 
$\widehat{\beta}_0$. Therefore a coupling slope large enough to reproduce the late-time evolution of $w_{\rm eff}$ cannot, in general, be extrapolated unchanged back to recombination. The local linear description makes this point transparent, i.e.\ the late-time phantom-like behavior is governed by the local quantities $(\alpha_0,\beta_0)$, whereas early-time viability is controlled by $\beta_{\rm rec}$. In general models these local coefficients need not coincide, and this separation provides a unified and model-independent way to discuss the late-time fit, the qualitative CMB requirement, and the existence of robust initial conditions.

\section{Realistic Example}
\label{sec:realization}
The local linear expansion analysis in Sec.~\ref{sec:evolution} already places strong restrictions on the allowed forms of the coupling between DE and DM, which implies that a broad class of commonly studied models may not simultaneously explain the DESI results and be CMB-safe, if starting with an approximately vanishing velocity in the RD era. For example, a previously studied model that does {\it not} satisfy our constraints, has the potential and coupling (e.g.~Refs.~\cite{Khoury:2003rn, Chakraborty:2025syu}) \footnote{The relative sign in the exponential of $f(\phi)$ is physically important, since it directly affects the direction of the field evolution and hence the evolution of the {\it effective} DE equation of state. In the present work we take $\lambda>0$, so that $f'(\phi)>0$ while $V'(\phi)<0$, consistent with the framework discussed in this paper where the slopes of the bare potential and the interaction function have opposite signs. In some other works, e.g.~Ref.~\cite{Li:2026xaz}, the interaction is instead taken in the form $f(\phi)\propto \exp[-|\lambda|\phi/(\sqrt{8\pi}M_{\rm Pl})]$, or equivalently on the $\lambda<0$ branch in the notation adopted in Eq.~\eqref{eq:exponential}. This corresponds to a qualitatively different scenario.}
\begin{align}
    V(\phi)  = V_0 \left( \frac{M_{\rm Pl}}{\phi}\right)^{\mu} \; , \quad 
     f(\phi)  = \exp\left(\frac{\lambda \phi}{\sqrt{8\pi} M_{\rm Pl}}\right) \; .
     \label{eq:exponential}
\end{align}
In this case, the effective coupling slope does not decrease toward earlier times.  As a result, if we require the evolution of $\phi$ to start from the frozen phase in the RD region, the choices of $\lambda$ that are large enough to generate the desired phantom-crossing behavior at $z\lesssim1$ may also imply an unacceptably strong interaction around recombination. Such models therefore tend to induce sizable modifications to the CMB when reproducing the low-redshift behavior motivated by DESI. In appendix~\ref{sec:exponential} we further study this class of models numerically in more detail. The result is consistent with the expectation from the local expansion analysis. Once the qualitative CMB requirements are imposed, the late-time evolution of $w_{\rm eff}$ becomes too mild to reproduce the phantom-crossing behavior at $z \lesssim 1$ motivated by DESI. In other words, within this setup the interaction can be made sufficiently weak near recombination only at the price of losing the strong late-time phantom-like evolution.

Motivated by the linear expansions of the previous section, 
we now consider a successful model which {\it does} satisfy our constraints. Concretely, we add a negative quadratic term to the linear coupling function, so as to suppress the slope of $f(\phi)$ at large field values. Without loss of generality, we shift the field such that the present-day value satisfies $\phi_0 = 0$. Then $V(\phi)$ and $f(\phi)$ can be respectively expressed for all times as
\begin{align}
V(\phi)=V_0+\alpha\phi \; , \quad 
\frac{f(\phi)}{f_0}=1+\beta\phi-\gamma\phi^2 \; .
\label{eq:quadratic}
\end{align}
Note that here $V(\phi)$ and $f(\phi)$ as defined in Eq.~(\ref{eq:quadratic}) are now taken to be the full functions over the field range of interest (not just approximate expansions) and $\alpha$, $\beta$ and $\gamma$ are universal coefficients which, once chosen, are valid at all epochs (times).\footnote{Note that $f(\phi)$ will become negative at very large field values for positive $\gamma$, so it does not represent an ultraviolet complete model.}
We restrict the choice of parameters to the branch $\alpha<0$, $\beta\ge 0$, $\gamma>0$. In addition, we also require $f(\phi)>0$ and $f'(\phi) = f_0(\beta - 2\gamma \phi)>0$ 
along the entire cosmological trajectory, which ensures a monotonic evolution toward $\phi\simeq 0$ today and avoids pathological regions where $f(\phi)$ changes sign.

For each point in parameter space, the background evolution is obtained by solving the coupled system for $(\phi,\dot\phi)$ in terms of $\ln a$, starting from $a = 10^{-7}$. The initial condition $\phi_{\rm ini}$ and the potential scale $V_0$ are not treated as free parameters, but are fixed by a shooting procedure requiring
\begin{align}
\phi_0 \simeq 0\; , \quad \rho_{\phi 0} \simeq \rho_{\rm DE0} \; ,
\label{eq:shooting}
\end{align}
so that the present-day Universe matches the observed DE abundance. 

\begin{figure}[t!]
        \centering
        \includegraphics[width=1\linewidth]{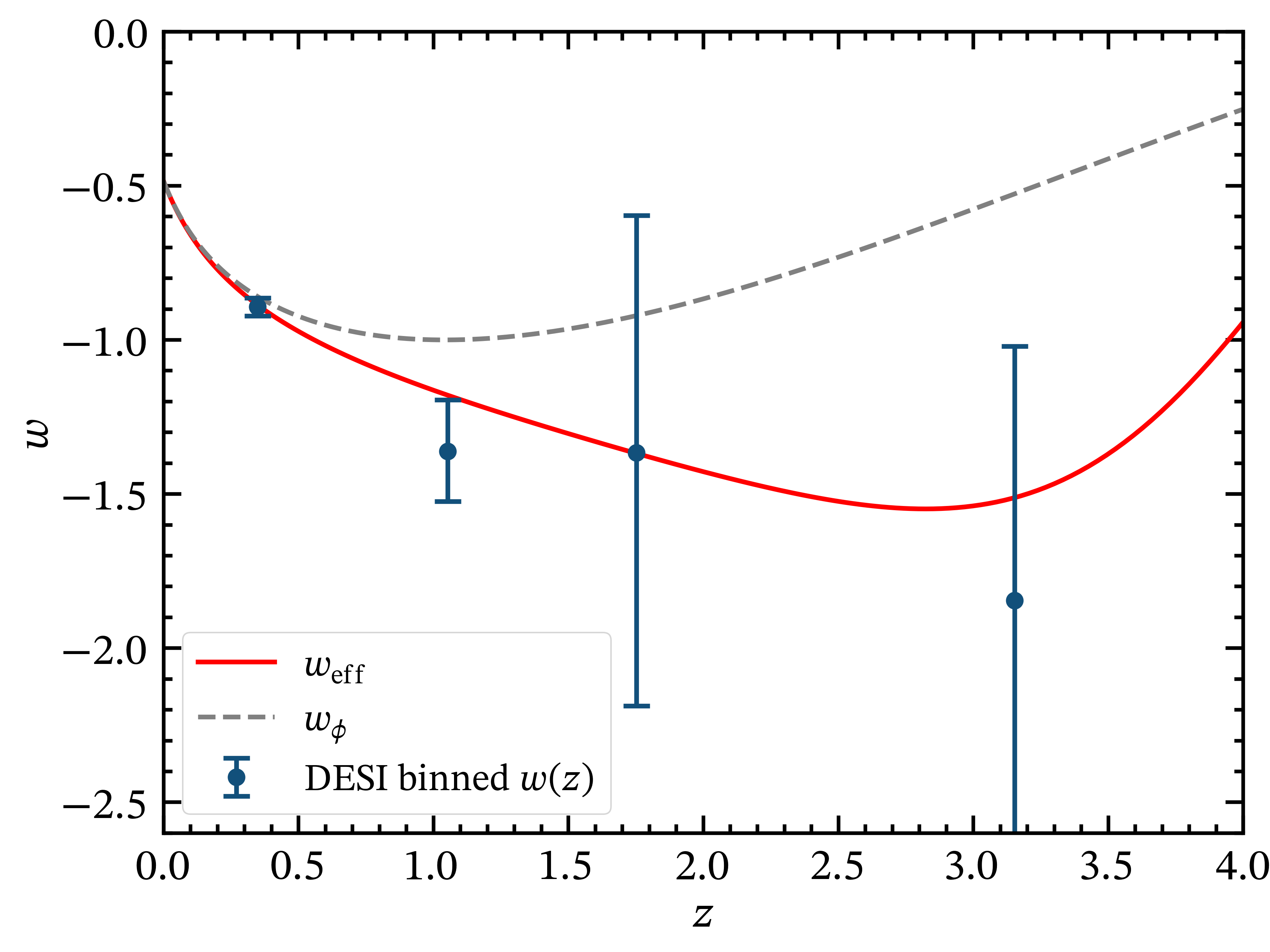}
        \caption{The evolution of $w(z)$ over the redshift $z$ in our realization with a linear bare potential and a quadratic interaction function shown in Eq.~\eqref{eq:quadratic}. The red curve shows the best-fit result of $w_{\rm eff}(z)$ where we choose $\alpha = -4.667\,H^2_0 M_{\rm Pl}$, $\beta = 0.180\,M^{-1}_{\rm Pl}$, $\gamma = 0.145 \, M_{\rm Pl}^{-2}$, and $\phi^\prime_{\rm ini} =0$ at $a_{\rm ini} = 10^{-7}$. The blue dashed curve corresponds to the evolution of $w_\phi(z)$. For comparison, we also show the $1\sigma$ error bars from the binned phenomenological reconstruction of $w(z)$ reported in the DESI DR2 extended DE analysis using DESI+CMB+Union3 data~\cite{DESI:2025fii, DESI:2025zgx}. All bins are fitted at the $1\sigma$ level. }
        \label{fig:best-fit}
\end{figure}

Given the solution, we construct the {\it effective} equation of state $w_{\rm eff}(z)$ using Eq.~\eqref{eq:weff}. The resulting $w_{\rm eff}(z)$ is then compared with the binned phenomenological reconstruction of $w(z)$ reported in the DESI DR2 extended DE analysis using DESI+CMB+Union3 data~\cite{DESI:2025fii, DESI:2025zgx}, which we adopt here only as an illustrative benchmark for the desired late-time behavior. We define a simple benchmark $\chi^2$ by treating the quoted $1\sigma$ intervals as independent Gaussian errors. The scan is performed over a restricted region in $(\alpha,\beta,\gamma)$
\begin{align}
    \alpha & \in [-7,-1]\,H^2_0 M_{\rm Pl} \; , \nonumber \\
    \beta & \in [0.1, 0.5]\, M^{-1}_{\rm Pl} \; , \nonumber \\
    \gamma &  \in [0.1, 0.5]\, M^{-2}_{\rm Pl} \; , \nonumber
\end{align}
which is motivated by the local analytical estimates, and the best-fit point is identified by minimizing $\chi^2$.

As discussed above, in addition to the late-time fit, we impose early-time viability conditions near recombination. In practice, we require
\begin{align}
\epsilon(z_{\rm rec})  < 0.005 \, , \;
|\Delta \ln  m|_{\rm rec}  < 0.01 \, , \;
\Omega_\phi(z_{\rm rec})   < 0.0036 \, . \nonumber
\end{align}
These conditions are implemented as hard cutoffs on the numerical solutions, ensuring that only trajectories consistent with the qualitative CMB requirements are retained. 

Finally, the Universe should still be undergoing accelerated expansion today, i.e.\ the deceleration parameter  $q \equiv -\ddot{a}/(a H^2)$ should be negative at the present time. In a spatially flat background, and neglecting the radiation contribution at the present epoch, this condition can be written as 
\begin{align}
    q_0 \simeq \frac12\left[\Omega_{\rm m}+\Omega_{\rm DE}\left(1+3w_{{\rm eff}}\right)\right]_{z=0}<0 \; ,
\end{align}
which leads to $w_{\rm eff}(z=0) \lesssim -0.48$ for our setup. We regard it as a minimal background-level consistency condition in the numerical analysis to ensure that the model does not contradict the observed late-time accelerated expansion of the Universe. The latter is supported most directly by the type-Ia supernova Hubble diagram~\cite{SupernovaCosmologyProject:1998vns,SupernovaSearchTeam:1998fmf}.

\section{Results and Discussion}
\label{sec:conclusion}

\begin{figure}[t!]
        \centering
        \includegraphics[width=1\linewidth]{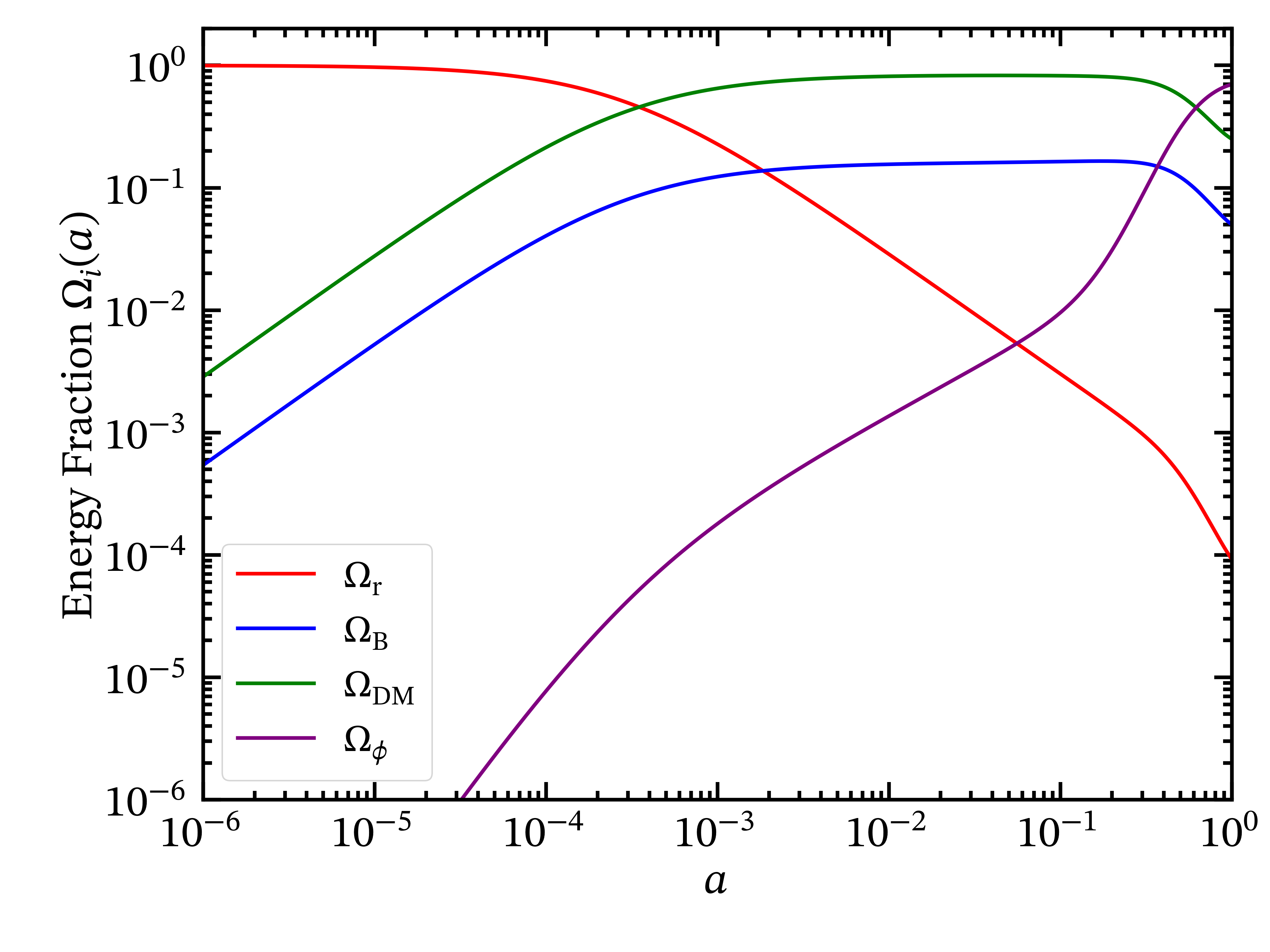}
        \caption{The evolution of energy fractions of radiation (red), baryons (blue), DM (green) and the scalar $\phi$ (purple) in terms of $a$, in our realization with a linear bare potential and a quadratic interaction function shown in Eq.~\eqref{eq:quadratic}. The parameter choice is the same as that in Fig.~\ref{fig:best-fit}.}
        \label{fig:fraction}
\end{figure}

\begin{figure}[t!]
        \centering
        \includegraphics[width=1\linewidth]{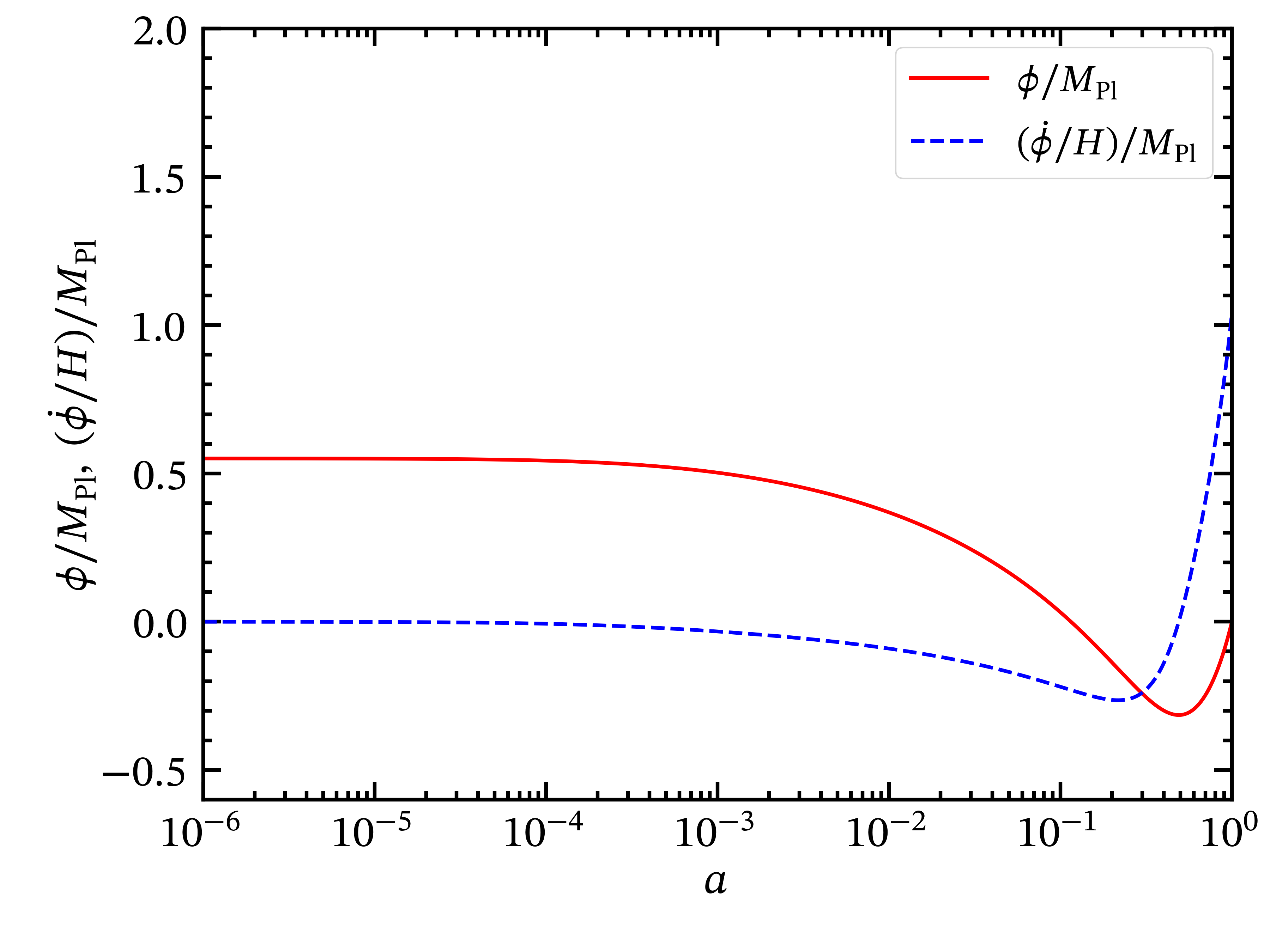}
        \caption{Red solid curve denotes the evolution of the field value $\phi$ in our realization with a linear bare potential and a quadratic interaction function shown in Eq.~\eqref{eq:quadratic}, while the blue dashed curve represents the evolution of $\dot\phi/H$ in terms of $a$. Both are normalized by $M_{\rm Pl}$. The parameter choice is the same as that in Fig.~\ref{fig:best-fit}.}
        \label{fig:field_evolution}
\end{figure}

Our results are shown in Figs.~\ref{fig:best-fit}--\ref{fig:field_evolution}. In Fig.~\ref{fig:best-fit}, we present the best-fit $w_{\rm eff}$ curve with $\chi^2 = 1.38$, corresponding to  $\alpha = -4.667\,H^2_0 M_{\rm Pl}$, $\beta = 0.180\,M^{-1}_{\rm Pl}$, $\gamma = 0.145 \, M_{\rm Pl}^{-2}$, and $\phi^\prime_{\rm ini} =0$ at $a_{\rm ini} = 10^{-7}$. For comparison, we also exhibit $1\sigma$ error bars from the binned phenomenological reconstruction of $w(z)$ in the DESI DR2 extended DE analysis~\cite{DESI:2025fii, DESI:2025zgx}. The resulting $w_{\rm eff}$ tracks the adopted binned-$w$ benchmark reasonably well. An {\it apparent} phantom crossing as shown by the parameter $w_{\rm eff}$ occurs within the range $0.5 \lesssim z \lesssim 1$, while the {\it bare} $w_\phi$ never drops below $-1$, as appropriate for a non-phantom field.

The evolution of energy fractions of different components is depicted in Fig.~\ref{fig:fraction}. By construction, the present-day energy fractions match the observed values through the shooting conditions in Eq.~\eqref{eq:shooting}. At earlier times, in particular around matter-radiation equality at $a \simeq 3\times 10^{-4}$, the scalar-field fraction $\Omega_\phi$ remains highly suppressed, $\Omega_\phi(a_{\rm rec}) \simeq 1.62 \times 10^{-4}$, and continues to stay at a negligible level until $a \simeq 10^{-1}$. This behavior indicates that the model recovers the standard RD and MD expansion history at the background level and to good approximation, while only allowing the DE sector to become dynamically relevant at sufficiently late times. 

We plot the evolution of the field value $\phi$ and its velocity $\phi'\equiv\dot\phi/H$ in Fig.~\ref{fig:field_evolution}, which indicates that $\phi$ initially remains nearly stationary and stays in an approximately frozen state until $a\simeq 10^{-3}$, and then begins to move toward smaller values of $\phi$, before reversing its direction of motion at a time close to the present epoch. This evolution is precisely consistent with the schematic picture illustrated in Fig.~\ref{fig:schematic}. At the recombination epoch $a_{\rm rec} \simeq 9.16\times 10^{-4}$, we obtain $\phi'(a_{\rm rec}) \simeq -3.12\times 10^{-2}\,M_{\rm Pl}$, $\epsilon(a_{\rm rec}) \simeq 9.78 \times 10^{-4}$ and $|\Delta \ln m|_{\rm rec} \simeq 7.01\times 10^{-4}$, all of which lie below the corresponding conservative CMB-safe bounds adopted in this work.

\section{Summary and Conclusion}
\label{conclusion}
In summary, we have considered an interacting DE scenario in which the DM mass is modulated by a Yukawa-type coupling to a quintessence field. Our main focus has been on the regime where the dark-sector interaction contributes a significant effective potential to the scalar field and therefore reshapes the DE evolution.

We have shown that a viable realization of this mechanism requires the scalar field to originate from a frozen phase deep in the radiation era, in order for the effective coupling to remain sufficiently suppressed before recombination to evade CMB constraints, and for the late-time evolution to become strong enough to reproduce the {\it apparent} behavior of $w_{\rm eff}(z)$ preferred by DESI.

We have proposed local linear expansions of the potential and coupling functions, and have identified a set of general requirements for simultaneously reproducing the low-redshift behavior of the inferred equation-of-state parameter and remaining qualitatively consistent with CMB constraints around recombination in a model-independent way. We have shown in particular that, if one aims for a robust frozen initial condition for the scalar field, the initial condition should be imposed deep in the RD era. We have further argued that CMB safety requires the energy transfer between DM and DE to remain sufficiently small near recombination, implying that the linear coefficient of the coupling function $f(\phi)$ around recombination must be small. By contrast, in order to account for the DESI-preferred {\it apparent} phantom-crossing behavior at low redshift, the corresponding local slope of the same $f(\phi)$ around the present epoch must be significantly larger. These requirements can be regarded as a guide for building future particle physics motivated DE models.

We have provided a realistic phenomenological example in which $w_{\rm eff}(z)$ evolves from $w_{\rm eff}\approx -1.2$ at $z \approx 1.0$ to $w_{\rm eff}\approx -0.9$ at $z\approx 0.4$. 
The resulting $w_{\rm eff}(z)$ can match reasonably well the DESI binned reconstruction within the $1\sigma$ error bars at low redshifts, while also satisfying conservative requirements of CMB safety around recombination. 

The present work constitutes a semi-quantitative analysis of the general viability of this framework, rather than as a full perturbation-level likelihood analysis against cosmological data. In particular, we have not yet included the evolution of perturbations in a Boltzmann treatment or confronted the model with precision cosmological observables in a full numerical fit. These issues are essential for assessing the ultimate viability of the scenario and are therefore left for future work.

Furthermore, the realistic example considered does not constitute a fully ultraviolet-complete construction. We have not attempted to build a complete model with a scalar potential manifestly bounded from below and an interaction structure guaranteed to avoid unphysical regions such as negative potentials over the full field range. Instead, it serves to illustrate an existence proof of the realization of the general conditions and requirements which serve as a guide for designing future models of this kind which can safely navigate the phantom divide.

{\bf Note added.}
At the final stage of preparing this manuscript, Ref.~\cite{Wang:2026wrk} appeared, which considered a model with a sign-switching interaction.

\acknowledgments
We are grateful to Prolay Chanda, Guo-Hong Du, William Giar{\`e} and Tian-Nuo Li for useful discussions.
SFK acknowledges the STFC Consolidated Grant
ST/X000583/1 and thanks IFIC, Valencia, for hospitality;
his work was funded by a Leverhulme Trust Emeritus Fellowship Grant.
XW is funded by the European Union, NextGenerationEU, National Recovery and Resilience Plan (mission 4, component 2)
under the project {\it MODIPAC: Modular Invariance in Particle Physics and Cosmology} (CUP C93C24004940006).  

\appendix

\section{Derivation of the general solution to the linearized field equation}
\label{app:derivation}
In this appendix, we derive the general solution to the linearized field equation. In the RD era, we have
\begin{align}
\phi''+\phi'+A_{\rm ini} a+B_{\rm ini} a^4\simeq 0\;.
\end{align}
To solve this equation, it is convenient to regard it as a first-order differential equation for
\begin{align}
y(N)\equiv \phi'(N)\;,
\end{align}
where $N\equiv \ln a$. Since $a=e^N$, the equation becomes
\begin{align}
y'+y+A_{\rm ini} e^N+B_{\rm ini} e^{4N}=0\;.
\end{align}
Multiplying both sides by $e^N$, we find
\begin{align}
e^N y'+e^N y
=
-A_{\rm ini} e^{2N}-B_{\rm ini} e^{5N}\;.
\end{align}
The left-hand side can be written as a total derivative,
\begin{align}
\frac{d}{dN}\left(e^N y\right)
=
-A_{\rm ini} e^{2N}-B_{\rm ini} e^{5N}\;.
\end{align}
Integrating once with respect to $N$, one obtains
\begin{align}
e^N y
=
-\frac{A_{\rm ini}}{2}e^{2N}
-\frac{B_{\rm ini}}{5}e^{5N}
+C\;,
\end{align}
where $C$ is an integration constant. Dividing by $e^N=a$ and restoring $y=\phi'$, we arrive at
\begin{align}
\phi'_{\rm RD}(a)
=
C a^{-1}
-\frac{A_{\rm ini}}{2}a
-\frac{B_{\rm ini}}{5}a^4\; ,
\end{align}
which is exactly the form shown in Eq.~\eqref{eq:RDgeneral}. The radiation-era attractor is obtained by setting $C=0$,
\begin{align}
\phi'_{\rm att,RD}(a)
=
-\frac{A_{\rm ini}}{2}a
-\frac{B_{\rm ini}}{5}a^4\;.
\end{align}

For the linearized field equation in the MD era, namely,
\begin{align}
\phi''+\frac{3}{2}\phi' +A_{\rm ini}^\prime +B_{\rm ini}^\prime \,a^3 \simeq 0\;,
\end{align}
we can follow a similar procedure and arrive at
\begin{align}
\phi'_{\rm MD}(a)
=
C'a^{-3/2}
-\frac{2}{3}A^\prime_{\rm ini}
-\frac{2}{9}B^\prime_{\rm ini}a^3\;.
\end{align}

\section{Numerical Study of the Exponential Coupling Model}
\label{sec:exponential}

In this appendix, we present the numerical results for the model with an inverse-power-law potential and an exponential coupling function, i.e.
\begin{align}
    V(\phi)  = V_0 \left( \frac{M_{\rm Pl}}{\phi}\right)^{\mu}, \quad 
     f(\phi)  = \exp\left(\frac{\lambda \phi}{\sqrt{8\pi} M_{\rm Pl}}\right) .
     \label{eq:exppotential}
\end{align}

\begin{figure}[t!]
        \centering
        \includegraphics[width=1\linewidth]{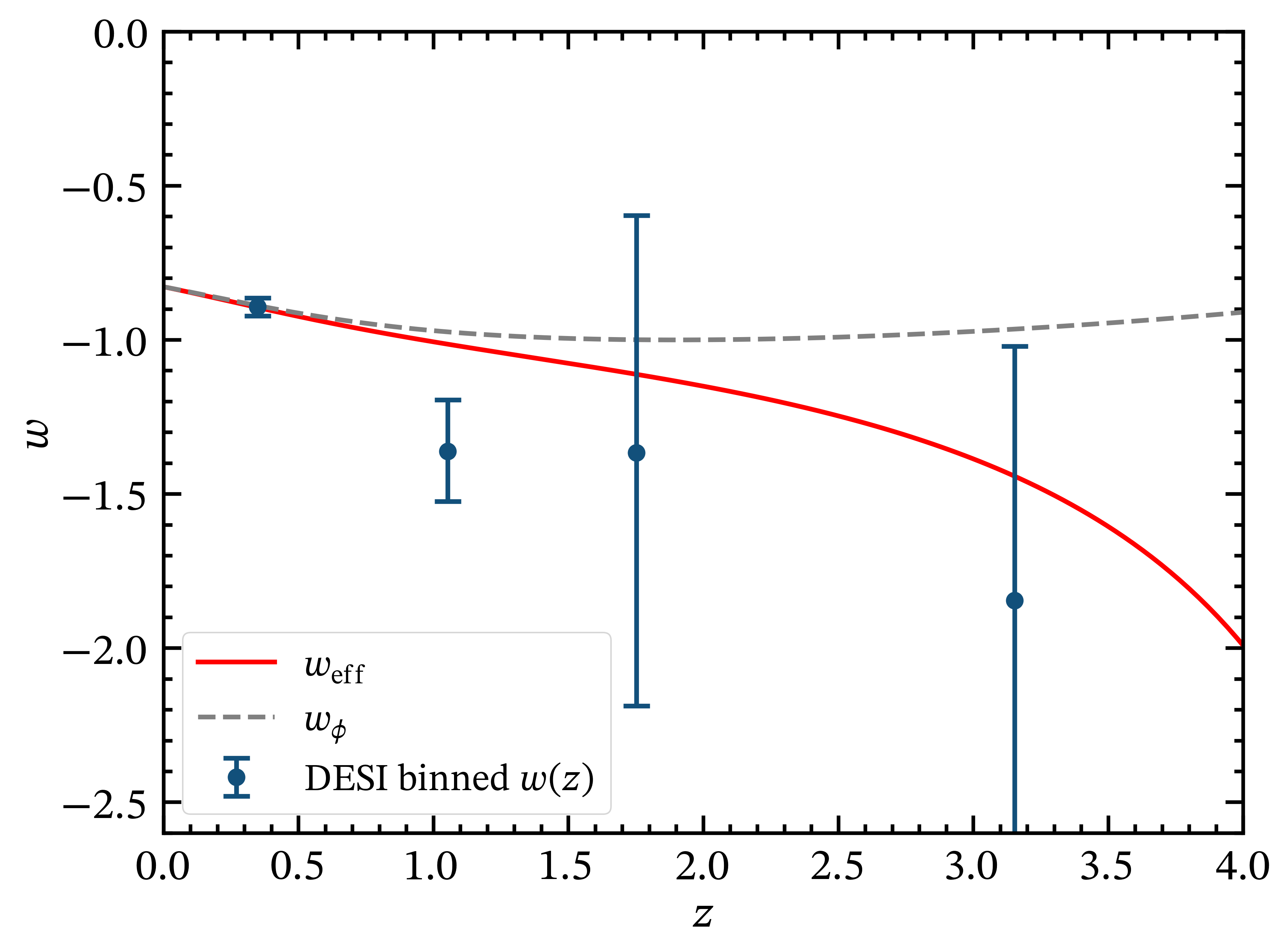}
        \caption{The evolution of $w(z)$ over the redshift $z$ in the model with an inverse-power-law potential and an exponential interacting function shown in  Eq.~\eqref{eq:exppotential}. The red curve shows the best-fit result of $w_{\rm eff}(z)$ where we choose $\mu = 1.75$, $\lambda = 0.30$, $f_0 = 1.11$, and $\phi^\prime_{\rm ini} =0$ at $a_{\rm ini} = 10^{-7}$. The blue dashed curve corresponds to the evolution of $w_\phi(z)$. For comparison, we also show the $1\sigma$ error bars from the binned phenomenological reconstruction of $w(z)$ reported in the DESI DR2 extended DE analysis using DESI+CMB+Union3 data~\cite{DESI:2025fii, DESI:2025zgx}. One can see that it is not possible to accommodate well the second bin.
        }
        \label{fig:best-fit-exp}
\end{figure}

The numerical strategy is analogous to that adopted in the main text for the quadratic interaction model, i.e.\ scanning the model parameters while imposing the late-time shooting conditions. The main difference is that, for the exponential coupling, the present-day field value is parameterized through
\begin{align}
    f_0 \equiv f(\phi_0) = \exp\!\left(\frac{\lambda \phi_0}{\sqrt{8\pi} M_{\rm Pl}}\right) \; ,
\end{align}
so that the shooting target for the present field value is adjusted to
\begin{align}
    \phi_0 = \frac{\sqrt{8\pi}\,M_{\rm Pl}}{\lambda}\ln f_0 \; .
\end{align}
For each point in parameter space, we then shoot on the initial field value and the overall scale $V_0$ such that the solution reproduces both the desired present-day field value and the observed DE density today. In the numerical scan we take
\begin{align}
    \mu \in [0.1,2.0]\;, \quad \lambda \in [0.1,0.7]\;, \quad f_0 \in (1.0,1.4]\; . \nonumber
\end{align}
As in the quadratic interaction model, the solutions are initialized deep in the RD era $a_{\rm ini} = 10^{-7}$ with a frozen-field initial condition $\phi'_{\rm ini} = 0$, and only those points that admit the CMB-safe conditions and a consistent late-time shooting solution are retained.

The best-fit point in this scan is found at
\begin{align}
    \mu = 1.75\;, \quad
    \lambda = 0.30\;, \quad
    f_0 = 1.11\; , \nonumber
\end{align}
with $\chi^2 \simeq 4.67$. The corresponding evolution of $w(z)$ is shown in Fig.~\ref{fig:best-fit-exp}. Although this fit still captures the qualitative behavior of an {\it effective} phantom crossing, the best-fit $w_{\rm eff}(z)$ curve shows a sizable deviation from the DESI-preferred bin around $z\sim 1$. Correspondingly, the {\it effective} phantom crossing is shifted to a higher redshift, taking place at $z\sim 1.5$. This indicates that in the exponential coupling model the local coupling slope around recombination cannot be efficiently separated from that at late times, so that satisfying the conservative early-time bounds tends to weaken the desired low-redshift phantom-like evolution, which is consistent with the general arguments in Sec.~\ref{sec:evolution}.

\bibliography{refs}

\begin{thebibliography}{81}%
\makeatletter
\providecommand \@ifxundefined [1]{%
 \@ifx{#1\undefined}
}%
\providecommand \@ifnum [1]{%
 \ifnum #1\expandafter \@firstoftwo
 \else \expandafter \@secondoftwo
 \fi
}%
\providecommand \@ifx [1]{%
 \ifx #1\expandafter \@firstoftwo
 \else \expandafter \@secondoftwo
 \fi
}%
\providecommand \natexlab [1]{#1}%
\providecommand \enquote  [1]{``#1''}%
\providecommand \bibnamefont  [1]{#1}%
\providecommand \bibfnamefont [1]{#1}%
\providecommand \citenamefont [1]{#1}%
\providecommand \href@noop [0]{\@secondoftwo}%
\providecommand \href [0]{\begingroup \@sanitize@url \@href}%
\providecommand \@href[1]{\@@startlink{#1}\@@href}%
\providecommand \@@href[1]{\endgroup#1\@@endlink}%
\providecommand \@sanitize@url [0]{\catcode `\\12\catcode `\$12\catcode
  `\&12\catcode `\#12\catcode `\^12\catcode `\_12\catcode `\%12\relax}%
\providecommand \@@startlink[1]{}%
\providecommand \@@endlink[0]{}%
\providecommand \url  [0]{\begingroup\@sanitize@url \@url }%
\providecommand \@url [1]{\endgroup\@href {#1}{\urlprefix }}%
\providecommand \urlprefix  [0]{URL }%
\providecommand \Eprint [0]{\href }%
\providecommand \doibase [0]{http://dx.doi.org/}%
\providecommand \selectlanguage [0]{\@gobble}%
\providecommand \bibinfo  [0]{\@secondoftwo}%
\providecommand \bibfield  [0]{\@secondoftwo}%
\providecommand \translation [1]{[#1]}%
\providecommand \BibitemOpen [0]{}%
\providecommand \bibitemStop [0]{}%
\providecommand \bibitemNoStop [0]{.\EOS\space}%
\providecommand \EOS [0]{\spacefactor3000\relax}%
\providecommand \BibitemShut  [1]{\csname bibitem#1\endcsname}%
\let\auto@bib@innerbib\@empty
\bibitem [{\citenamefont {Efstathiou}\ \emph {et~al.}(1990)\citenamefont
  {Efstathiou}, \citenamefont {Sutherland},\ and\ \citenamefont
  {Maddox}}]{Efstathiou:1990xe}%
  \BibitemOpen
  \bibfield  {author} {\bibinfo {author} {\bibfnamefont {G.}~\bibnamefont
  {Efstathiou}}, \bibinfo {author} {\bibfnamefont {W.~J.}\ \bibnamefont
  {Sutherland}}, \ and\ \bibinfo {author} {\bibfnamefont {S.~J.}\ \bibnamefont
  {Maddox}},\ }\bibfield  {title} {\enquote {\bibinfo {title} {{The
  cosmological constant and cold dark matter}},}\ }\href {\doibase
  10.1038/348705a0} {\bibfield  {journal} {\bibinfo  {journal} {Nature}\
  }\textbf {\bibinfo {volume} {348}},\ \bibinfo {pages} {705--707} (\bibinfo
  {year} {1990})}\BibitemShut {NoStop}%
\bibitem [{\citenamefont {Frieman}\ \emph {et~al.}(2008)\citenamefont
  {Frieman}, \citenamefont {Turner},\ and\ \citenamefont
  {Huterer}}]{Frieman:2008sn}%
  \BibitemOpen
  \bibfield  {author} {\bibinfo {author} {\bibfnamefont {Joshua}\ \bibnamefont
  {Frieman}}, \bibinfo {author} {\bibfnamefont {Michael}\ \bibnamefont
  {Turner}}, \ and\ \bibinfo {author} {\bibfnamefont {Dragan}\ \bibnamefont
  {Huterer}},\ }\bibfield  {title} {\enquote {\bibinfo {title} {{Dark Energy
  and the Accelerating Universe}},}\ }\href {\doibase
  10.1146/annurev.astro.46.060407.145243} {\bibfield  {journal} {\bibinfo
  {journal} {Ann. Rev. Astron. Astrophys.}\ }\textbf {\bibinfo {volume} {46}},\
  \bibinfo {pages} {385--432} (\bibinfo {year} {2008})},\ \Eprint
  {http://arxiv.org/abs/0803.0982} {arXiv:0803.0982 [astro-ph]} \BibitemShut
  {NoStop}%
\bibitem [{\citenamefont {Weinberg}\ \emph {et~al.}(2013)\citenamefont
  {Weinberg}, \citenamefont {Mortonson}, \citenamefont {Eisenstein},
  \citenamefont {Hirata}, \citenamefont {Riess},\ and\ \citenamefont
  {Rozo}}]{Weinberg:2013agg}%
  \BibitemOpen
  \bibfield  {author} {\bibinfo {author} {\bibfnamefont {David~H.}\
  \bibnamefont {Weinberg}}, \bibinfo {author} {\bibfnamefont {Michael~J.}\
  \bibnamefont {Mortonson}}, \bibinfo {author} {\bibfnamefont {Daniel~J.}\
  \bibnamefont {Eisenstein}}, \bibinfo {author} {\bibfnamefont {Christopher}\
  \bibnamefont {Hirata}}, \bibinfo {author} {\bibfnamefont {Adam~G.}\
  \bibnamefont {Riess}}, \ and\ \bibinfo {author} {\bibfnamefont {Eduardo}\
  \bibnamefont {Rozo}},\ }\bibfield  {title} {\enquote {\bibinfo {title}
  {{Observational Probes of Cosmic Acceleration}},}\ }\href {\doibase
  10.1016/j.physrep.2013.05.001} {\bibfield  {journal} {\bibinfo  {journal}
  {Phys. Rept.}\ }\textbf {\bibinfo {volume} {530}},\ \bibinfo {pages}
  {87--255} (\bibinfo {year} {2013})},\ \Eprint
  {http://arxiv.org/abs/1201.2434} {arXiv:1201.2434 [astro-ph.CO]} \BibitemShut
  {NoStop}%
\bibitem [{\citenamefont {Riess}\ \emph {et~al.}(1998)\citenamefont {Riess}
  \emph {et~al.}}]{SupernovaSearchTeam:1998fmf}%
  \BibitemOpen
  \bibfield  {author} {\bibinfo {author} {\bibfnamefont {Adam~G.}\ \bibnamefont
  {Riess}} \emph {et~al.} (\bibinfo {collaboration} {Supernova Search Team}),\
  }\bibfield  {title} {\enquote {\bibinfo {title} {{Observational evidence from
  supernovae for an accelerating universe and a cosmological constant}},}\
  }\href {\doibase 10.1086/300499} {\bibfield  {journal} {\bibinfo  {journal}
  {Astron. J.}\ }\textbf {\bibinfo {volume} {116}},\ \bibinfo {pages}
  {1009--1038} (\bibinfo {year} {1998})},\ \Eprint
  {http://arxiv.org/abs/astro-ph/9805201} {arXiv:astro-ph/9805201} \BibitemShut
  {NoStop}%
\bibitem [{\citenamefont {Perlmutter}\ \emph {et~al.}(1999)\citenamefont
  {Perlmutter} \emph {et~al.}}]{SupernovaCosmologyProject:1998vns}%
  \BibitemOpen
  \bibfield  {author} {\bibinfo {author} {\bibfnamefont {S.}~\bibnamefont
  {Perlmutter}} \emph {et~al.} (\bibinfo {collaboration} {Supernova Cosmology
  Project}),\ }\bibfield  {title} {\enquote {\bibinfo {title} {{Measurements of
  $\Omega$ and $\Lambda$ from 42 High Redshift Supernovae}},}\ }\href {\doibase
  10.1086/307221} {\bibfield  {journal} {\bibinfo  {journal} {Astrophys. J.}\
  }\textbf {\bibinfo {volume} {517}},\ \bibinfo {pages} {565--586} (\bibinfo
  {year} {1999})},\ \Eprint {http://arxiv.org/abs/astro-ph/9812133}
  {arXiv:astro-ph/9812133} \BibitemShut {NoStop}%
\bibitem [{\citenamefont {Percival}\ \emph {et~al.}(2002)\citenamefont
  {Percival} \emph {et~al.}}]{2dFGRSTeam:2002tzq}%
  \BibitemOpen
  \bibfield  {author} {\bibinfo {author} {\bibfnamefont {Will~J.}\ \bibnamefont
  {Percival}} \emph {et~al.} (\bibinfo {collaboration} {2dFGRS Team}),\
  }\bibfield  {title} {\enquote {\bibinfo {title} {{Parameter constraints for
  flat cosmologies from CMB and 2dFGRS power spectra}},}\ }\href {\doibase
  10.1046/j.1365-8711.2002.06001.x} {\bibfield  {journal} {\bibinfo  {journal}
  {Mon. Not. Roy. Astron. Soc.}\ }\textbf {\bibinfo {volume} {337}},\ \bibinfo
  {pages} {1068} (\bibinfo {year} {2002})},\ \Eprint
  {http://arxiv.org/abs/astro-ph/0206256} {arXiv:astro-ph/0206256} \BibitemShut
  {NoStop}%
\bibitem [{\citenamefont {Cole}\ \emph {et~al.}(2005)\citenamefont {Cole} \emph
  {et~al.}}]{2dFGRS:2005yhx}%
  \BibitemOpen
  \bibfield  {author} {\bibinfo {author} {\bibfnamefont {Shaun}\ \bibnamefont
  {Cole}} \emph {et~al.} (\bibinfo {collaboration} {2dFGRS}),\ }\bibfield
  {title} {\enquote {\bibinfo {title} {{The 2dF Galaxy Redshift Survey:
  Power-spectrum analysis of the final dataset and cosmological
  implications}},}\ }\href {\doibase 10.1111/j.1365-2966.2005.09318.x}
  {\bibfield  {journal} {\bibinfo  {journal} {Mon. Not. Roy. Astron. Soc.}\
  }\textbf {\bibinfo {volume} {362}},\ \bibinfo {pages} {505--534} (\bibinfo
  {year} {2005})},\ \Eprint {http://arxiv.org/abs/astro-ph/0501174}
  {arXiv:astro-ph/0501174} \BibitemShut {NoStop}%
\bibitem [{\citenamefont {Aghanim}\ \emph {et~al.}(2020)\citenamefont {Aghanim}
  \emph {et~al.}}]{Planck:2018vyg}%
  \BibitemOpen
  \bibfield  {author} {\bibinfo {author} {\bibfnamefont {N.}~\bibnamefont
  {Aghanim}} \emph {et~al.} (\bibinfo {collaboration} {Planck}),\ }\bibfield
  {title} {\enquote {\bibinfo {title} {{Planck 2018 results. VI. Cosmological
  parameters}},}\ }\href {\doibase 10.1051/0004-6361/201833910} {\bibfield
  {journal} {\bibinfo  {journal} {Astron. Astrophys.}\ }\textbf {\bibinfo
  {volume} {641}},\ \bibinfo {pages} {A6} (\bibinfo {year} {2020})},\ \bibinfo
  {note} {[Erratum: Astron.Astrophys. 652, C4 (2021)]},\ \Eprint
  {http://arxiv.org/abs/1807.06209} {arXiv:1807.06209 [astro-ph.CO]}
  \BibitemShut {NoStop}%
\bibitem [{\citenamefont {Zhao}\ \emph {et~al.}(2022)\citenamefont {Zhao} \emph
  {et~al.}}]{eBOSS:2021pff}%
  \BibitemOpen
  \bibfield  {author} {\bibinfo {author} {\bibfnamefont {Cheng}\ \bibnamefont
  {Zhao}} \emph {et~al.} (\bibinfo {collaboration} {eBOSS}),\ }\bibfield
  {title} {\enquote {\bibinfo {title} {{The completed SDSS-IV extended Baryon
  Oscillation Spectroscopic Survey: cosmological implications from multitracer
  BAO analysis with galaxies and voids}},}\ }\href {\doibase
  10.1093/mnras/stac390} {\bibfield  {journal} {\bibinfo  {journal} {Mon. Not.
  Roy. Astron. Soc.}\ }\textbf {\bibinfo {volume} {511}},\ \bibinfo {pages}
  {5492--5524} (\bibinfo {year} {2022})},\ \Eprint
  {http://arxiv.org/abs/2110.03824} {arXiv:2110.03824 [astro-ph.CO]}
  \BibitemShut {NoStop}%
\bibitem [{\citenamefont {Abbott}\ \emph {et~al.}(2018)\citenamefont {Abbott}
  \emph {et~al.}}]{DES:2017myr}%
  \BibitemOpen
  \bibfield  {author} {\bibinfo {author} {\bibfnamefont {T.~M.~C.}\
  \bibnamefont {Abbott}} \emph {et~al.} (\bibinfo {collaboration} {DES}),\
  }\bibfield  {title} {\enquote {\bibinfo {title} {{Dark Energy Survey year 1
  results: Cosmological constraints from galaxy clustering and weak
  lensing}},}\ }\href {\doibase 10.1103/PhysRevD.98.043526} {\bibfield
  {journal} {\bibinfo  {journal} {Phys. Rev. D}\ }\textbf {\bibinfo {volume}
  {98}},\ \bibinfo {pages} {043526} (\bibinfo {year} {2018})},\ \Eprint
  {http://arxiv.org/abs/1708.01530} {arXiv:1708.01530 [astro-ph.CO]}
  \BibitemShut {NoStop}%
\bibitem [{\citenamefont {Alam}\ \emph {et~al.}(2021)\citenamefont {Alam} \emph
  {et~al.}}]{eBOSS:2020yzd}%
  \BibitemOpen
  \bibfield  {author} {\bibinfo {author} {\bibfnamefont {Shadab}\ \bibnamefont
  {Alam}} \emph {et~al.} (\bibinfo {collaboration} {eBOSS}),\ }\bibfield
  {title} {\enquote {\bibinfo {title} {{Completed SDSS-IV extended Baryon
  Oscillation Spectroscopic Survey: Cosmological implications from two decades
  of spectroscopic surveys at the Apache Point Observatory}},}\ }\href
  {\doibase 10.1103/PhysRevD.103.083533} {\bibfield  {journal} {\bibinfo
  {journal} {Phys. Rev. D}\ }\textbf {\bibinfo {volume} {103}},\ \bibinfo
  {pages} {083533} (\bibinfo {year} {2021})},\ \Eprint
  {http://arxiv.org/abs/2007.08991} {arXiv:2007.08991 [astro-ph.CO]}
  \BibitemShut {NoStop}%
\bibitem [{\citenamefont {Heymans}\ \emph {et~al.}(2021)\citenamefont {Heymans}
  \emph {et~al.}}]{Heymans:2020gsg}%
  \BibitemOpen
  \bibfield  {author} {\bibinfo {author} {\bibfnamefont {Catherine}\
  \bibnamefont {Heymans}} \emph {et~al.},\ }\bibfield  {title} {\enquote
  {\bibinfo {title} {{KiDS-1000 Cosmology: Multi-probe weak gravitational
  lensing and spectroscopic galaxy clustering constraints}},}\ }\href {\doibase
  10.1051/0004-6361/202039063} {\bibfield  {journal} {\bibinfo  {journal}
  {Astron. Astrophys.}\ }\textbf {\bibinfo {volume} {646}},\ \bibinfo {pages}
  {A140} (\bibinfo {year} {2021})},\ \Eprint {http://arxiv.org/abs/2007.15632}
  {arXiv:2007.15632 [astro-ph.CO]} \BibitemShut {NoStop}%
\bibitem [{\citenamefont {Brout}\ \emph {et~al.}(2022)\citenamefont {Brout}
  \emph {et~al.}}]{Brout:2022vxf}%
  \BibitemOpen
  \bibfield  {author} {\bibinfo {author} {\bibfnamefont {Dillon}\ \bibnamefont
  {Brout}} \emph {et~al.},\ }\bibfield  {title} {\enquote {\bibinfo {title}
  {{The Pantheon+ Analysis: Cosmological Constraints}},}\ }\href {\doibase
  10.3847/1538-4357/ac8e04} {\bibfield  {journal} {\bibinfo  {journal}
  {Astrophys. J.}\ }\textbf {\bibinfo {volume} {938}},\ \bibinfo {pages} {110}
  (\bibinfo {year} {2022})},\ \Eprint {http://arxiv.org/abs/2202.04077}
  {arXiv:2202.04077 [astro-ph.CO]} \BibitemShut {NoStop}%
\bibitem [{\citenamefont {Abbott}\ \emph {et~al.}(2024)\citenamefont {Abbott}
  \emph {et~al.}}]{DES:2024jxu}%
  \BibitemOpen
  \bibfield  {author} {\bibinfo {author} {\bibfnamefont {T.~M.~C.}\
  \bibnamefont {Abbott}} \emph {et~al.} (\bibinfo {collaboration} {DES}),\
  }\bibfield  {title} {\enquote {\bibinfo {title} {{The Dark Energy Survey:
  Cosmology Results with {\ensuremath{\sim}}1500 New High-redshift Type Ia
  Supernovae Using the Full 5 yr Data Set}},}\ }\href {\doibase
  10.3847/2041-8213/ad6f9f} {\bibfield  {journal} {\bibinfo  {journal}
  {Astrophys. J. Lett.}\ }\textbf {\bibinfo {volume} {973}},\ \bibinfo {pages}
  {L14} (\bibinfo {year} {2024})},\ \Eprint {http://arxiv.org/abs/2401.02929}
  {arXiv:2401.02929 [astro-ph.CO]} \BibitemShut {NoStop}%
\bibitem [{\citenamefont {Abdalla}\ \emph {et~al.}(2022)\citenamefont {Abdalla}
  \emph {et~al.}}]{Abdalla:2022yfr}%
  \BibitemOpen
  \bibfield  {author} {\bibinfo {author} {\bibfnamefont {Elcio}\ \bibnamefont
  {Abdalla}} \emph {et~al.},\ }\bibfield  {title} {\enquote {\bibinfo {title}
  {{Cosmology intertwined: A review of the particle physics, astrophysics, and
  cosmology associated with the cosmological tensions and anomalies}},}\ }\href
  {\doibase 10.1016/j.jheap.2022.04.002} {\bibfield  {journal} {\bibinfo
  {journal} {JHEAp}\ }\textbf {\bibinfo {volume} {34}},\ \bibinfo {pages}
  {49--211} (\bibinfo {year} {2022})},\ \Eprint
  {http://arxiv.org/abs/2203.06142} {arXiv:2203.06142 [astro-ph.CO]}
  \BibitemShut {NoStop}%
\bibitem [{\citenamefont {Perivolaropoulos}\ and\ \citenamefont
  {Skara}(2022)}]{Perivolaropoulos:2021jda}%
  \BibitemOpen
  \bibfield  {author} {\bibinfo {author} {\bibfnamefont {Leandros}\
  \bibnamefont {Perivolaropoulos}}\ and\ \bibinfo {author} {\bibfnamefont
  {Foteini}\ \bibnamefont {Skara}},\ }\bibfield  {title} {\enquote {\bibinfo
  {title} {{Challenges for {\ensuremath{\Lambda}}CDM: An update}},}\ }\href
  {\doibase 10.1016/j.newar.2022.101659} {\bibfield  {journal} {\bibinfo
  {journal} {New Astron. Rev.}\ }\textbf {\bibinfo {volume} {95}},\ \bibinfo
  {pages} {101659} (\bibinfo {year} {2022})},\ \Eprint
  {http://arxiv.org/abs/2105.05208} {arXiv:2105.05208 [astro-ph.CO]}
  \BibitemShut {NoStop}%
\bibitem [{\citenamefont {Adame}\ \emph {et~al.}(2025)\citenamefont {Adame}
  \emph {et~al.}}]{DESI:2024mwx}%
  \BibitemOpen
  \bibfield  {author} {\bibinfo {author} {\bibfnamefont {A.~G.}\ \bibnamefont
  {Adame}} \emph {et~al.} (\bibinfo {collaboration} {DESI}),\ }\bibfield
  {title} {\enquote {\bibinfo {title} {{DESI 2024 VI: cosmological constraints
  from the measurements of baryon acoustic oscillations}},}\ }\href {\doibase
  10.1088/1475-7516/2025/02/021} {\bibfield  {journal} {\bibinfo  {journal}
  {JCAP}\ }\textbf {\bibinfo {volume} {02}},\ \bibinfo {pages} {021} (\bibinfo
  {year} {2025})},\ \Eprint {http://arxiv.org/abs/2404.03002} {arXiv:2404.03002
  [astro-ph.CO]} \BibitemShut {NoStop}%
\bibitem [{\citenamefont {Lodha}\ \emph {et~al.}(2025)\citenamefont {Lodha}
  \emph {et~al.}}]{DESI:2025fii}%
  \BibitemOpen
  \bibfield  {author} {\bibinfo {author} {\bibfnamefont {K.}~\bibnamefont
  {Lodha}} \emph {et~al.} (\bibinfo {collaboration} {DESI}),\ }\bibfield
  {title} {\enquote {\bibinfo {title} {{Extended dark energy analysis using
  DESI DR2 BAO measurements}},}\ }\href {\doibase 10.1103/w4c6-1r5j} {\bibfield
   {journal} {\bibinfo  {journal} {Phys. Rev. D}\ }\textbf {\bibinfo {volume}
  {112}},\ \bibinfo {pages} {083511} (\bibinfo {year} {2025})},\ \Eprint
  {http://arxiv.org/abs/2503.14743} {arXiv:2503.14743 [astro-ph.CO]}
  \BibitemShut {NoStop}%
\bibitem [{\citenamefont {Abdul~Karim}\ \emph {et~al.}(2025)\citenamefont
  {Abdul~Karim} \emph {et~al.}}]{DESI:2025zgx}%
  \BibitemOpen
  \bibfield  {author} {\bibinfo {author} {\bibfnamefont {M.}~\bibnamefont
  {Abdul~Karim}} \emph {et~al.} (\bibinfo {collaboration} {DESI}),\ }\bibfield
  {title} {\enquote {\bibinfo {title} {{DESI DR2 results. II. Measurements of
  baryon acoustic oscillations and cosmological constraints}},}\ }\href
  {\doibase 10.1103/tr6y-kpc6} {\bibfield  {journal} {\bibinfo  {journal}
  {Phys. Rev. D}\ }\textbf {\bibinfo {volume} {112}},\ \bibinfo {pages}
  {083515} (\bibinfo {year} {2025})},\ \Eprint
  {http://arxiv.org/abs/2503.14738} {arXiv:2503.14738 [astro-ph.CO]}
  \BibitemShut {NoStop}%
\bibitem [{\citenamefont {Andrade}\ \emph {et~al.}(2025)\citenamefont {Andrade}
  \emph {et~al.}}]{DESI:2025qqy}%
  \BibitemOpen
  \bibfield  {author} {\bibinfo {author} {\bibfnamefont {U.}~\bibnamefont
  {Andrade}} \emph {et~al.} (\bibinfo {collaboration} {DESI}),\ }\bibfield
  {title} {\enquote {\bibinfo {title} {{Validation of the DESI DR2 measurements
  of baryon acoustic oscillations from galaxies and quasars}},}\ }\href
  {\doibase 10.1103/kdys-w8vl} {\bibfield  {journal} {\bibinfo  {journal}
  {Phys. Rev. D}\ }\textbf {\bibinfo {volume} {112}},\ \bibinfo {pages}
  {083512} (\bibinfo {year} {2025})},\ \Eprint
  {http://arxiv.org/abs/2503.14742} {arXiv:2503.14742 [astro-ph.CO]}
  \BibitemShut {NoStop}%
\bibitem [{\citenamefont {Elbers}\ \emph {et~al.}(2025)\citenamefont {Elbers}
  \emph {et~al.}}]{Elbers:2025vlz}%
  \BibitemOpen
  \bibfield  {author} {\bibinfo {author} {\bibfnamefont {W.}~\bibnamefont
  {Elbers}} \emph {et~al.},\ }\bibfield  {title} {\enquote {\bibinfo {title}
  {{Constraints on neutrino physics from DESI DR2 BAO and DR1 full shape}},}\
  }\href {\doibase 10.1103/w9pk-xsk7} {\bibfield  {journal} {\bibinfo
  {journal} {Phys. Rev. D}\ }\textbf {\bibinfo {volume} {112}},\ \bibinfo
  {pages} {083513} (\bibinfo {year} {2025})},\ \Eprint
  {http://arxiv.org/abs/2503.14744} {arXiv:2503.14744 [astro-ph.CO]}
  \BibitemShut {NoStop}%
\bibitem [{\citenamefont {Lee}\ \emph {et~al.}(2022)\citenamefont {Lee},
  \citenamefont {Lee}, \citenamefont {Colg{\'a}in}, \citenamefont
  {Sheikh-Jabbari},\ and\ \citenamefont {Thakur}}]{Lee:2022cyh}%
  \BibitemOpen
  \bibfield  {author} {\bibinfo {author} {\bibfnamefont {Bum-Hoon}\
  \bibnamefont {Lee}}, \bibinfo {author} {\bibfnamefont {Wonwoo}\ \bibnamefont
  {Lee}}, \bibinfo {author} {\bibfnamefont {Eoin~{\'O}.}\ \bibnamefont
  {Colg{\'a}in}}, \bibinfo {author} {\bibfnamefont {M.~M.}\ \bibnamefont
  {Sheikh-Jabbari}}, \ and\ \bibinfo {author} {\bibfnamefont {Somyadip}\
  \bibnamefont {Thakur}},\ }\bibfield  {title} {\enquote {\bibinfo {title} {{Is
  local H $_{0}$ at odds with dark energy EFT?}}}\ }\href {\doibase
  10.1088/1475-7516/2022/04/004} {\bibfield  {journal} {\bibinfo  {journal}
  {JCAP}\ }\textbf {\bibinfo {volume} {04}},\ \bibinfo {pages} {004} (\bibinfo
  {year} {2022})},\ \Eprint {http://arxiv.org/abs/2202.03906} {arXiv:2202.03906
  [astro-ph.CO]} \BibitemShut {NoStop}%
\bibitem [{\citenamefont {Colg{\'a}in}\ \emph {et~al.}(2026)\citenamefont
  {Colg{\'a}in}, \citenamefont {Pourojaghi}, \citenamefont {Sheikh-Jabbari},\
  and\ \citenamefont {Yin}}]{Colgain:2025nzf}%
  \BibitemOpen
  \bibfield  {author} {\bibinfo {author} {\bibfnamefont {Eoin~{\'O}.}\
  \bibnamefont {Colg{\'a}in}}, \bibinfo {author} {\bibfnamefont {Saeed}\
  \bibnamefont {Pourojaghi}}, \bibinfo {author} {\bibfnamefont {M.~M.}\
  \bibnamefont {Sheikh-Jabbari}}, \ and\ \bibinfo {author} {\bibfnamefont
  {Lu}~\bibnamefont {Yin}},\ }\bibfield  {title} {\enquote {\bibinfo {title}
  {{How much has DESI dark energy evolved since DR1?}}}\ }\href {\doibase
  10.1016/j.dark.2026.102268} {\bibfield  {journal} {\bibinfo  {journal} {Phys.
  Dark Univ.}\ }\textbf {\bibinfo {volume} {52}},\ \bibinfo {pages} {102268}
  (\bibinfo {year} {2026})},\ \Eprint {http://arxiv.org/abs/2504.04417}
  {arXiv:2504.04417 [astro-ph.CO]} \BibitemShut {NoStop}%
\bibitem [{\citenamefont {Wetterich}(1995)}]{Wetterich:1994bg}%
  \BibitemOpen
  \bibfield  {author} {\bibinfo {author} {\bibfnamefont {Christof}\
  \bibnamefont {Wetterich}},\ }\bibfield  {title} {\enquote {\bibinfo {title}
  {{The Cosmon model for an asymptotically vanishing time dependent
  cosmological 'constant'}},}\ }\href@noop {} {\bibfield  {journal} {\bibinfo
  {journal} {Astron. Astrophys.}\ }\textbf {\bibinfo {volume} {301}},\ \bibinfo
  {pages} {321--328} (\bibinfo {year} {1995})},\ \Eprint
  {http://arxiv.org/abs/hep-th/9408025} {arXiv:hep-th/9408025} \BibitemShut
  {NoStop}%
\bibitem [{\citenamefont {Ratra}\ and\ \citenamefont
  {Peebles}(1988)}]{Ratra:1987rm}%
  \BibitemOpen
  \bibfield  {author} {\bibinfo {author} {\bibfnamefont {Bharat}\ \bibnamefont
  {Ratra}}\ and\ \bibinfo {author} {\bibfnamefont {P.~J.~E.}\ \bibnamefont
  {Peebles}},\ }\bibfield  {title} {\enquote {\bibinfo {title} {{Cosmological
  Consequences of a Rolling Homogeneous Scalar Field}},}\ }\href {\doibase
  10.1103/PhysRevD.37.3406} {\bibfield  {journal} {\bibinfo  {journal} {Phys.
  Rev. D}\ }\textbf {\bibinfo {volume} {37}},\ \bibinfo {pages} {3406}
  (\bibinfo {year} {1988})}\BibitemShut {NoStop}%
\bibitem [{\citenamefont {Sahni}\ and\ \citenamefont
  {Starobinsky}(2000)}]{Sahni:1999gb}%
  \BibitemOpen
  \bibfield  {author} {\bibinfo {author} {\bibfnamefont {Varun}\ \bibnamefont
  {Sahni}}\ and\ \bibinfo {author} {\bibfnamefont {Alexei~A.}\ \bibnamefont
  {Starobinsky}},\ }\bibfield  {title} {\enquote {\bibinfo {title} {{The Case
  for a positive cosmological Lambda term}},}\ }\href {\doibase
  10.1142/S0218271800000542} {\bibfield  {journal} {\bibinfo  {journal} {Int.
  J. Mod. Phys. D}\ }\textbf {\bibinfo {volume} {9}},\ \bibinfo {pages}
  {373--444} (\bibinfo {year} {2000})},\ \Eprint
  {http://arxiv.org/abs/astro-ph/9904398} {arXiv:astro-ph/9904398} \BibitemShut
  {NoStop}%
\bibitem [{\citenamefont {Peebles}\ and\ \citenamefont
  {Ratra}(2003)}]{Peebles:2002gy}%
  \BibitemOpen
  \bibfield  {author} {\bibinfo {author} {\bibfnamefont {P.~J.~E.}\
  \bibnamefont {Peebles}}\ and\ \bibinfo {author} {\bibfnamefont {Bharat}\
  \bibnamefont {Ratra}},\ }\bibfield  {title} {\enquote {\bibinfo {title} {{The
  Cosmological Constant and Dark Energy}},}\ }\href {\doibase
  10.1103/RevModPhys.75.559} {\bibfield  {journal} {\bibinfo  {journal} {Rev.
  Mod. Phys.}\ }\textbf {\bibinfo {volume} {75}},\ \bibinfo {pages} {559--606}
  (\bibinfo {year} {2003})},\ \Eprint {http://arxiv.org/abs/astro-ph/0207347}
  {arXiv:astro-ph/0207347} \BibitemShut {NoStop}%
\bibitem [{\citenamefont {Guo}\ and\ \citenamefont {Zhang}(2005)}]{Guo:2004vg}%
  \BibitemOpen
  \bibfield  {author} {\bibinfo {author} {\bibfnamefont {Zong-Kuan}\
  \bibnamefont {Guo}}\ and\ \bibinfo {author} {\bibfnamefont {Yuan-Zhong}\
  \bibnamefont {Zhang}},\ }\bibfield  {title} {\enquote {\bibinfo {title}
  {{Interacting phantom energy}},}\ }\href {\doibase
  10.1103/PhysRevD.71.023501} {\bibfield  {journal} {\bibinfo  {journal} {Phys.
  Rev. D}\ }\textbf {\bibinfo {volume} {71}},\ \bibinfo {pages} {023501}
  (\bibinfo {year} {2005})},\ \Eprint {http://arxiv.org/abs/astro-ph/0411524}
  {arXiv:astro-ph/0411524} \BibitemShut {NoStop}%
\bibitem [{\citenamefont {Amendola}(2000)}]{Amendola:1999er}%
  \BibitemOpen
  \bibfield  {author} {\bibinfo {author} {\bibfnamefont {Luca}\ \bibnamefont
  {Amendola}},\ }\bibfield  {title} {\enquote {\bibinfo {title} {{Coupled
  quintessence}},}\ }\href {\doibase 10.1103/PhysRevD.62.043511} {\bibfield
  {journal} {\bibinfo  {journal} {Phys. Rev. D}\ }\textbf {\bibinfo {volume}
  {62}},\ \bibinfo {pages} {043511} (\bibinfo {year} {2000})},\ \Eprint
  {http://arxiv.org/abs/astro-ph/9908023} {arXiv:astro-ph/9908023} \BibitemShut
  {NoStop}%
\bibitem [{\citenamefont {Farrar}\ and\ \citenamefont
  {Peebles}(2004)}]{Farrar:2003uw}%
  \BibitemOpen
  \bibfield  {author} {\bibinfo {author} {\bibfnamefont {Glennys~R.}\
  \bibnamefont {Farrar}}\ and\ \bibinfo {author} {\bibfnamefont {P.~James~E.}\
  \bibnamefont {Peebles}},\ }\bibfield  {title} {\enquote {\bibinfo {title}
  {{Interacting dark matter and dark energy}},}\ }\href {\doibase
  10.1086/381728} {\bibfield  {journal} {\bibinfo  {journal} {Astrophys. J.}\
  }\textbf {\bibinfo {volume} {604}},\ \bibinfo {pages} {1--11} (\bibinfo
  {year} {2004})},\ \Eprint {http://arxiv.org/abs/astro-ph/0307316}
  {arXiv:astro-ph/0307316} \BibitemShut {NoStop}%
\bibitem [{\citenamefont {Khoury}\ and\ \citenamefont
  {Weltman}(2004)}]{Khoury:2003rn}%
  \BibitemOpen
  \bibfield  {author} {\bibinfo {author} {\bibfnamefont {Justin}\ \bibnamefont
  {Khoury}}\ and\ \bibinfo {author} {\bibfnamefont {Amanda}\ \bibnamefont
  {Weltman}},\ }\bibfield  {title} {\enquote {\bibinfo {title} {{Chameleon
  cosmology}},}\ }\href {\doibase 10.1103/PhysRevD.69.044026} {\bibfield
  {journal} {\bibinfo  {journal} {Phys. Rev. D}\ }\textbf {\bibinfo {volume}
  {69}},\ \bibinfo {pages} {044026} (\bibinfo {year} {2004})},\ \Eprint
  {http://arxiv.org/abs/astro-ph/0309411} {arXiv:astro-ph/0309411} \BibitemShut
  {NoStop}%
\bibitem [{\citenamefont {Das}\ \emph {et~al.}(2006)\citenamefont {Das},
  \citenamefont {Corasaniti},\ and\ \citenamefont {Khoury}}]{Das:2005yj}%
  \BibitemOpen
  \bibfield  {author} {\bibinfo {author} {\bibfnamefont {Subinoy}\ \bibnamefont
  {Das}}, \bibinfo {author} {\bibfnamefont {Pier~Stefano}\ \bibnamefont
  {Corasaniti}}, \ and\ \bibinfo {author} {\bibfnamefont {Justin}\ \bibnamefont
  {Khoury}},\ }\bibfield  {title} {\enquote {\bibinfo {title}
  {{Super-acceleration as signature of dark sector interaction}},}\ }\href
  {\doibase 10.1103/PhysRevD.73.083509} {\bibfield  {journal} {\bibinfo
  {journal} {Phys. Rev. D}\ }\textbf {\bibinfo {volume} {73}},\ \bibinfo
  {pages} {083509} (\bibinfo {year} {2006})},\ \Eprint
  {http://arxiv.org/abs/astro-ph/0510628} {arXiv:astro-ph/0510628} \BibitemShut
  {NoStop}%
\bibitem [{\citenamefont {Antusch}\ \emph {et~al.}(2008)\citenamefont
  {Antusch}, \citenamefont {Das},\ and\ \citenamefont
  {Dutta}}]{Antusch:2008hj}%
  \BibitemOpen
  \bibfield  {author} {\bibinfo {author} {\bibfnamefont {Stefan}\ \bibnamefont
  {Antusch}}, \bibinfo {author} {\bibfnamefont {Subinoy}\ \bibnamefont {Das}},
  \ and\ \bibinfo {author} {\bibfnamefont {Koushik}\ \bibnamefont {Dutta}},\
  }\bibfield  {title} {\enquote {\bibinfo {title} {{Phenomenology of Hybrid
  Scenarios of Neutrino Dark Energy}},}\ }\href {\doibase
  10.1088/1475-7516/2008/10/016} {\bibfield  {journal} {\bibinfo  {journal}
  {JCAP}\ }\textbf {\bibinfo {volume} {10}},\ \bibinfo {pages} {016} (\bibinfo
  {year} {2008})},\ \Eprint {http://arxiv.org/abs/0807.4930} {arXiv:0807.4930
  [astro-ph]} \BibitemShut {NoStop}%
\bibitem [{\citenamefont {Cai}\ \emph {et~al.}(2021)\citenamefont {Cai},
  \citenamefont {Guo}, \citenamefont {Li}, \citenamefont {Wang},\ and\
  \citenamefont {Yu}}]{Cai:2021wgv}%
  \BibitemOpen
  \bibfield  {author} {\bibinfo {author} {\bibfnamefont {Rong-Gen}\
  \bibnamefont {Cai}}, \bibinfo {author} {\bibfnamefont {Zong-Kuan}\
  \bibnamefont {Guo}}, \bibinfo {author} {\bibfnamefont {Li}~\bibnamefont
  {Li}}, \bibinfo {author} {\bibfnamefont {Shao-Jiang}\ \bibnamefont {Wang}}, \
  and\ \bibinfo {author} {\bibfnamefont {Wang-Wei}\ \bibnamefont {Yu}},\
  }\bibfield  {title} {\enquote {\bibinfo {title} {{Chameleon dark energy can
  resolve the Hubble tension}},}\ }\href {\doibase
  10.1103/PhysRevD.103.L121302} {\bibfield  {journal} {\bibinfo  {journal}
  {Phys. Rev. D}\ }\textbf {\bibinfo {volume} {103}},\ \bibinfo {pages}
  {121302} (\bibinfo {year} {2021})},\ \Eprint
  {http://arxiv.org/abs/2102.02020} {arXiv:2102.02020 [astro-ph.CO]}
  \BibitemShut {NoStop}%
\bibitem [{\citenamefont {Copeland}\ \emph {et~al.}(2006)\citenamefont
  {Copeland}, \citenamefont {Sami},\ and\ \citenamefont
  {Tsujikawa}}]{Copeland:2006wr}%
  \BibitemOpen
  \bibfield  {author} {\bibinfo {author} {\bibfnamefont {Edmund~J.}\
  \bibnamefont {Copeland}}, \bibinfo {author} {\bibfnamefont {M.}~\bibnamefont
  {Sami}}, \ and\ \bibinfo {author} {\bibfnamefont {Shinji}\ \bibnamefont
  {Tsujikawa}},\ }\bibfield  {title} {\enquote {\bibinfo {title} {{Dynamics of
  dark energy}},}\ }\href {\doibase 10.1142/S021827180600942X} {\bibfield
  {journal} {\bibinfo  {journal} {Int. J. Mod. Phys. D}\ }\textbf {\bibinfo
  {volume} {15}},\ \bibinfo {pages} {1753--1936} (\bibinfo {year} {2006})},\
  \Eprint {http://arxiv.org/abs/hep-th/0603057} {arXiv:hep-th/0603057}
  \BibitemShut {NoStop}%
\bibitem [{\citenamefont {Guo}\ \emph {et~al.}(2007)\citenamefont {Guo},
  \citenamefont {Ohta},\ and\ \citenamefont {Tsujikawa}}]{Guo:2007zk}%
  \BibitemOpen
  \bibfield  {author} {\bibinfo {author} {\bibfnamefont {Zong-Kuan}\
  \bibnamefont {Guo}}, \bibinfo {author} {\bibfnamefont {Nobuyoshi}\
  \bibnamefont {Ohta}}, \ and\ \bibinfo {author} {\bibfnamefont {Shinji}\
  \bibnamefont {Tsujikawa}},\ }\bibfield  {title} {\enquote {\bibinfo {title}
  {{Probing the Coupling between Dark Components of the Universe}},}\ }\href
  {\doibase 10.1103/PhysRevD.76.023508} {\bibfield  {journal} {\bibinfo
  {journal} {Phys. Rev. D}\ }\textbf {\bibinfo {volume} {76}},\ \bibinfo
  {pages} {023508} (\bibinfo {year} {2007})},\ \Eprint
  {http://arxiv.org/abs/astro-ph/0702015} {arXiv:astro-ph/0702015} \BibitemShut
  {NoStop}%
\bibitem [{\citenamefont {Cai}\ and\ \citenamefont {Su}(2010)}]{Cai:2009ht}%
  \BibitemOpen
  \bibfield  {author} {\bibinfo {author} {\bibfnamefont {Rong-Gen}\
  \bibnamefont {Cai}}\ and\ \bibinfo {author} {\bibfnamefont {Qiping}\
  \bibnamefont {Su}},\ }\bibfield  {title} {\enquote {\bibinfo {title} {{On the
  Dark Sector Interactions}},}\ }\href {\doibase 10.1103/PhysRevD.81.103514}
  {\bibfield  {journal} {\bibinfo  {journal} {Phys. Rev. D}\ }\textbf {\bibinfo
  {volume} {81}},\ \bibinfo {pages} {103514} (\bibinfo {year} {2010})},\
  \Eprint {http://arxiv.org/abs/0912.1943} {arXiv:0912.1943 [astro-ph.CO]}
  \BibitemShut {NoStop}%
\bibitem [{\citenamefont {Bull}\ \emph {et~al.}(2016)\citenamefont {Bull} \emph
  {et~al.}}]{Bull:2015stt}%
  \BibitemOpen
  \bibfield  {author} {\bibinfo {author} {\bibfnamefont {Philip}\ \bibnamefont
  {Bull}} \emph {et~al.},\ }\bibfield  {title} {\enquote {\bibinfo {title}
  {{Beyond $\Lambda$CDM: Problems, solutions, and the road ahead}},}\ }\href
  {\doibase 10.1016/j.dark.2016.02.001} {\bibfield  {journal} {\bibinfo
  {journal} {Phys. Dark Univ.}\ }\textbf {\bibinfo {volume} {12}},\ \bibinfo
  {pages} {56--99} (\bibinfo {year} {2016})},\ \Eprint
  {http://arxiv.org/abs/1512.05356} {arXiv:1512.05356 [astro-ph.CO]}
  \BibitemShut {NoStop}%
\bibitem [{\citenamefont {Zlatev}\ \emph {et~al.}(1999)\citenamefont {Zlatev},
  \citenamefont {Wang},\ and\ \citenamefont {Steinhardt}}]{Zlatev:1998tr}%
  \BibitemOpen
  \bibfield  {author} {\bibinfo {author} {\bibfnamefont {Ivaylo}\ \bibnamefont
  {Zlatev}}, \bibinfo {author} {\bibfnamefont {Li-Min}\ \bibnamefont {Wang}}, \
  and\ \bibinfo {author} {\bibfnamefont {Paul~J.}\ \bibnamefont {Steinhardt}},\
  }\bibfield  {title} {\enquote {\bibinfo {title} {{Quintessence, cosmic
  coincidence, and the cosmological constant}},}\ }\href {\doibase
  10.1103/PhysRevLett.82.896} {\bibfield  {journal} {\bibinfo  {journal} {Phys.
  Rev. Lett.}\ }\textbf {\bibinfo {volume} {82}},\ \bibinfo {pages} {896--899}
  (\bibinfo {year} {1999})},\ \Eprint {http://arxiv.org/abs/astro-ph/9807002}
  {arXiv:astro-ph/9807002} \BibitemShut {NoStop}%
\bibitem [{\citenamefont {Caldwell}\ and\ \citenamefont
  {Linder}(2005)}]{Caldwell:2005tm}%
  \BibitemOpen
  \bibfield  {author} {\bibinfo {author} {\bibfnamefont {R.~R.}\ \bibnamefont
  {Caldwell}}\ and\ \bibinfo {author} {\bibfnamefont {Eric~V.}\ \bibnamefont
  {Linder}},\ }\bibfield  {title} {\enquote {\bibinfo {title} {{The Limits of
  quintessence}},}\ }\href {\doibase 10.1103/PhysRevLett.95.141301} {\bibfield
  {journal} {\bibinfo  {journal} {Phys. Rev. Lett.}\ }\textbf {\bibinfo
  {volume} {95}},\ \bibinfo {pages} {141301} (\bibinfo {year} {2005})},\
  \Eprint {http://arxiv.org/abs/astro-ph/0505494} {arXiv:astro-ph/0505494}
  \BibitemShut {NoStop}%
\bibitem [{\citenamefont {Caldwell}(2002)}]{Caldwell:1999ew}%
  \BibitemOpen
  \bibfield  {author} {\bibinfo {author} {\bibfnamefont {R.~R.}\ \bibnamefont
  {Caldwell}},\ }\bibfield  {title} {\enquote {\bibinfo {title} {{A Phantom
  menace?}}}\ }\href {\doibase 10.1016/S0370-2693(02)02589-3} {\bibfield
  {journal} {\bibinfo  {journal} {Phys. Lett. B}\ }\textbf {\bibinfo {volume}
  {545}},\ \bibinfo {pages} {23--29} (\bibinfo {year} {2002})},\ \Eprint
  {http://arxiv.org/abs/astro-ph/9908168} {arXiv:astro-ph/9908168} \BibitemShut
  {NoStop}%
\bibitem [{\citenamefont {Cline}\ \emph {et~al.}(2004)\citenamefont {Cline},
  \citenamefont {Jeon},\ and\ \citenamefont {Moore}}]{Cline:2003gs}%
  \BibitemOpen
  \bibfield  {author} {\bibinfo {author} {\bibfnamefont {James~M.}\
  \bibnamefont {Cline}}, \bibinfo {author} {\bibfnamefont {Sangyong}\
  \bibnamefont {Jeon}}, \ and\ \bibinfo {author} {\bibfnamefont {Guy~D.}\
  \bibnamefont {Moore}},\ }\bibfield  {title} {\enquote {\bibinfo {title} {{The
  Phantom menaced: Constraints on low-energy effective ghosts}},}\ }\href
  {\doibase 10.1103/PhysRevD.70.043543} {\bibfield  {journal} {\bibinfo
  {journal} {Phys. Rev. D}\ }\textbf {\bibinfo {volume} {70}},\ \bibinfo
  {pages} {043543} (\bibinfo {year} {2004})},\ \Eprint
  {http://arxiv.org/abs/hep-ph/0311312} {arXiv:hep-ph/0311312} \BibitemShut
  {NoStop}%
\bibitem [{\citenamefont {Feng}\ \emph {et~al.}(2005)\citenamefont {Feng},
  \citenamefont {Wang},\ and\ \citenamefont {Zhang}}]{Feng:2004ad}%
  \BibitemOpen
  \bibfield  {author} {\bibinfo {author} {\bibfnamefont {Bo}~\bibnamefont
  {Feng}}, \bibinfo {author} {\bibfnamefont {Xiu-Lian}\ \bibnamefont {Wang}}, \
  and\ \bibinfo {author} {\bibfnamefont {Xin-Min}\ \bibnamefont {Zhang}},\
  }\bibfield  {title} {\enquote {\bibinfo {title} {{Dark energy constraints
  from the cosmic age and supernova}},}\ }\href {\doibase
  10.1016/j.physletb.2004.12.071} {\bibfield  {journal} {\bibinfo  {journal}
  {Phys. Lett. B}\ }\textbf {\bibinfo {volume} {607}},\ \bibinfo {pages}
  {35--41} (\bibinfo {year} {2005})},\ \Eprint
  {http://arxiv.org/abs/astro-ph/0404224} {arXiv:astro-ph/0404224} \BibitemShut
  {NoStop}%
\bibitem [{\citenamefont {Ludwick}(2017)}]{Ludwick:2017tox}%
  \BibitemOpen
  \bibfield  {author} {\bibinfo {author} {\bibfnamefont {Kevin~J.}\
  \bibnamefont {Ludwick}},\ }\bibfield  {title} {\enquote {\bibinfo {title}
  {{The viability of phantom dark energy: A review}},}\ }\href {\doibase
  10.1142/S0217732317300257} {\bibfield  {journal} {\bibinfo  {journal} {Mod.
  Phys. Lett. A}\ }\textbf {\bibinfo {volume} {32}},\ \bibinfo {pages}
  {1730025} (\bibinfo {year} {2017})},\ \Eprint
  {http://arxiv.org/abs/1708.06981} {arXiv:1708.06981 [astro-ph.CO]}
  \BibitemShut {NoStop}%
\bibitem [{\citenamefont {Cai}\ \emph {et~al.}(2025)\citenamefont {Cai},
  \citenamefont {Ren}, \citenamefont {Qiu}, \citenamefont {Li},\ and\
  \citenamefont {Zhang}}]{Cai:2025mas}%
  \BibitemOpen
  \bibfield  {author} {\bibinfo {author} {\bibfnamefont {Yifu}\ \bibnamefont
  {Cai}}, \bibinfo {author} {\bibfnamefont {Xin}\ \bibnamefont {Ren}}, \bibinfo
  {author} {\bibfnamefont {Taotao}\ \bibnamefont {Qiu}}, \bibinfo {author}
  {\bibfnamefont {Mingzhe}\ \bibnamefont {Li}}, \ and\ \bibinfo {author}
  {\bibfnamefont {Xinmin}\ \bibnamefont {Zhang}},\ }\bibfield  {title}
  {\enquote {\bibinfo {title} {{The Quintom theory of dark energy after DESI
  DR2}},}\ }\href {\doibase 10.1093/nsr/nwag115} {\  (\bibinfo {year} {2025}),\
  10.1093/nsr/nwag115},\ \Eprint {http://arxiv.org/abs/2505.24732}
  {arXiv:2505.24732 [astro-ph.CO]} \BibitemShut {NoStop}%
\bibitem [{\citenamefont {Giar{\`e}}\ \emph {et~al.}(2024)\citenamefont
  {Giar{\`e}}, \citenamefont {Najafi}, \citenamefont {Pan}, \citenamefont
  {Di~Valentino},\ and\ \citenamefont {Firouzjaee}}]{Giare:2024gpk}%
  \BibitemOpen
  \bibfield  {author} {\bibinfo {author} {\bibfnamefont {William}\ \bibnamefont
  {Giar{\`e}}}, \bibinfo {author} {\bibfnamefont {Mahdi}\ \bibnamefont
  {Najafi}}, \bibinfo {author} {\bibfnamefont {Supriya}\ \bibnamefont {Pan}},
  \bibinfo {author} {\bibfnamefont {Eleonora}\ \bibnamefont {Di~Valentino}}, \
  and\ \bibinfo {author} {\bibfnamefont {Javad~T.}\ \bibnamefont
  {Firouzjaee}},\ }\bibfield  {title} {\enquote {\bibinfo {title} {{Robust
  preference for Dynamical Dark Energy in DESI BAO and SN measurements}},}\
  }\href {\doibase 10.1088/1475-7516/2024/10/035} {\bibfield  {journal}
  {\bibinfo  {journal} {JCAP}\ }\textbf {\bibinfo {volume} {10}},\ \bibinfo
  {pages} {035} (\bibinfo {year} {2024})},\ \Eprint
  {http://arxiv.org/abs/2407.16689} {arXiv:2407.16689 [astro-ph.CO]}
  \BibitemShut {NoStop}%
\bibitem [{\citenamefont {Li}\ \emph {et~al.}(2024)\citenamefont {Li},
  \citenamefont {Wu}, \citenamefont {Du}, \citenamefont {Jin}, \citenamefont
  {Li}, \citenamefont {Zhang},\ and\ \citenamefont {Zhang}}]{Li:2024qso}%
  \BibitemOpen
  \bibfield  {author} {\bibinfo {author} {\bibfnamefont {Tian-Nuo}\
  \bibnamefont {Li}}, \bibinfo {author} {\bibfnamefont {Peng-Ju}\ \bibnamefont
  {Wu}}, \bibinfo {author} {\bibfnamefont {Guo-Hong}\ \bibnamefont {Du}},
  \bibinfo {author} {\bibfnamefont {Shang-Jie}\ \bibnamefont {Jin}}, \bibinfo
  {author} {\bibfnamefont {Hai-Li}\ \bibnamefont {Li}}, \bibinfo {author}
  {\bibfnamefont {Jing-Fei}\ \bibnamefont {Zhang}}, \ and\ \bibinfo {author}
  {\bibfnamefont {Xin}\ \bibnamefont {Zhang}},\ }\bibfield  {title} {\enquote
  {\bibinfo {title} {{Constraints on Interacting Dark Energy Models from the
  DESI Baryon Acoustic Oscillation and DES Supernovae Data}},}\ }\href
  {\doibase 10.3847/1538-4357/ad87f0} {\bibfield  {journal} {\bibinfo
  {journal} {Astrophys. J.}\ }\textbf {\bibinfo {volume} {976}},\ \bibinfo
  {pages} {1} (\bibinfo {year} {2024})},\ \Eprint
  {http://arxiv.org/abs/2407.14934} {arXiv:2407.14934 [astro-ph.CO]}
  \BibitemShut {NoStop}%
\bibitem [{\citenamefont {Sabogal}\ \emph {et~al.}(2025)\citenamefont
  {Sabogal}, \citenamefont {Silva}, \citenamefont {Nunes}, \citenamefont
  {Kumar},\ and\ \citenamefont {Di~Valentino}}]{Sabogal:2025mkp}%
  \BibitemOpen
  \bibfield  {author} {\bibinfo {author} {\bibfnamefont {Miguel~A.}\
  \bibnamefont {Sabogal}}, \bibinfo {author} {\bibfnamefont {Emanuelly}\
  \bibnamefont {Silva}}, \bibinfo {author} {\bibfnamefont {Rafael~C.}\
  \bibnamefont {Nunes}}, \bibinfo {author} {\bibfnamefont {Suresh}\
  \bibnamefont {Kumar}}, \ and\ \bibinfo {author} {\bibfnamefont {Eleonora}\
  \bibnamefont {Di~Valentino}},\ }\bibfield  {title} {\enquote {\bibinfo
  {title} {{Sign switching in dark sector coupling interactions as a candidate
  for resolving cosmological tensions}},}\ }\href {\doibase
  10.1103/PhysRevD.111.043531} {\bibfield  {journal} {\bibinfo  {journal}
  {Phys. Rev. D}\ }\textbf {\bibinfo {volume} {111}},\ \bibinfo {pages}
  {043531} (\bibinfo {year} {2025})},\ \Eprint
  {http://arxiv.org/abs/2501.10323} {arXiv:2501.10323 [astro-ph.CO]}
  \BibitemShut {NoStop}%
\bibitem [{\citenamefont {Wolf}\ \emph {et~al.}(2025)\citenamefont {Wolf},
  \citenamefont {Garc\'{\i}a-Garc\'{\i}a}, \citenamefont {Anton},\ and\
  \citenamefont {Ferreira}}]{Wolf:2025jed}%
  \BibitemOpen
  \bibfield  {author} {\bibinfo {author} {\bibfnamefont {William~J.}\
  \bibnamefont {Wolf}}, \bibinfo {author} {\bibfnamefont {Carlos}\ \bibnamefont
  {Garc\'{\i}a-Garc\'{\i}a}}, \bibinfo {author} {\bibfnamefont {Theodore}\
  \bibnamefont {Anton}}, \ and\ \bibinfo {author} {\bibfnamefont {Pedro~G.}\
  \bibnamefont {Ferreira}},\ }\href {\doibase 10.1103/PhysRevLett.135.081001}
  {\bibfield  {journal} {\bibinfo  {journal} {Phys. Rev. Lett.}\ }\textbf
  {\bibinfo {volume} {135}},\ \bibinfo {pages} {081001} (\bibinfo {year}
  {2025})},\ \Eprint {http://arxiv.org/abs/2504.07679} {arXiv:2504.07679
  [astro-ph.CO]} \BibitemShut {NoStop}%
\bibitem [{\citenamefont {Li}\ \emph {et~al.}(2026{\natexlab{a}})\citenamefont
  {Li}, \citenamefont {Du}, \citenamefont {Li}, \citenamefont {Wu},
  \citenamefont {Jin}, \citenamefont {Zhang},\ and\ \citenamefont
  {Zhang}}]{Li:2025owk}%
  \BibitemOpen
  \bibfield  {author} {\bibinfo {author} {\bibfnamefont {Tian-Nuo}\
  \bibnamefont {Li}}, \bibinfo {author} {\bibfnamefont {Guo-Hong}\ \bibnamefont
  {Du}}, \bibinfo {author} {\bibfnamefont {Yun-He}\ \bibnamefont {Li}},
  \bibinfo {author} {\bibfnamefont {Peng-Ju}\ \bibnamefont {Wu}}, \bibinfo
  {author} {\bibfnamefont {Shang-Jie}\ \bibnamefont {Jin}}, \bibinfo {author}
  {\bibfnamefont {Jing-Fei}\ \bibnamefont {Zhang}}, \ and\ \bibinfo {author}
  {\bibfnamefont {Xin}\ \bibnamefont {Zhang}},\ }\bibfield  {title} {\enquote
  {\bibinfo {title} {{Probing the sign-changeable interaction between dark
  energy and dark matter with DESI baryon acoustic oscillations and DES
  supernovae data}},}\ }\href {\doibase 10.1007/s11433-025-2771-5} {\bibfield
  {journal} {\bibinfo  {journal} {Sci. China Phys. Mech. Astron.}\ }\textbf
  {\bibinfo {volume} {69}},\ \bibinfo {pages} {210413} (\bibinfo {year}
  {2026}{\natexlab{a}})},\ \Eprint {http://arxiv.org/abs/2501.07361}
  {arXiv:2501.07361 [astro-ph.CO]} \BibitemShut {NoStop}%
\bibitem [{\citenamefont {Dinda}\ and\ \citenamefont
  {Maartens}(2025)}]{Dinda:2025iaq}%
  \BibitemOpen
  \bibfield  {author} {\bibinfo {author} {\bibfnamefont {Bikash~R.}\
  \bibnamefont {Dinda}}\ and\ \bibinfo {author} {\bibfnamefont {Roy}\
  \bibnamefont {Maartens}},\ }\bibfield  {title} {\enquote {\bibinfo {title}
  {{Physical vs phantom dark energy after DESI: thawing quintessence in a
  curved background}},}\ }\href {\doibase 10.1093/mnrasl/slaf063} {\bibfield
  {journal} {\bibinfo  {journal} {Mon. Not. Roy. Astron. Soc.}\ }\textbf
  {\bibinfo {volume} {542}},\ \bibinfo {pages} {L31--L35} (\bibinfo {year}
  {2025})},\ \Eprint {http://arxiv.org/abs/2504.15190} {arXiv:2504.15190
  [astro-ph.CO]} \BibitemShut {NoStop}%
\bibitem [{\citenamefont {de~Souza}\ \emph {et~al.}(2025)\citenamefont
  {de~Souza}, \citenamefont {Rodrigues},\ and\ \citenamefont
  {Alcaniz}}]{deSouza:2025rhv}%
  \BibitemOpen
  \bibfield  {author} {\bibinfo {author} {\bibfnamefont {Rayff}\ \bibnamefont
  {de~Souza}}, \bibinfo {author} {\bibfnamefont {Gabriel}\ \bibnamefont
  {Rodrigues}}, \ and\ \bibinfo {author} {\bibfnamefont {Jailson}\ \bibnamefont
  {Alcaniz}},\ }\bibfield  {title} {\enquote {\bibinfo {title} {{Thawing
  quintessence and transient cosmic acceleration in light of DESI}},}\ }\href
  {\doibase 10.1103/2tjq-dtbc} {\bibfield  {journal} {\bibinfo  {journal}
  {Phys. Rev. D}\ }\textbf {\bibinfo {volume} {112}},\ \bibinfo {pages}
  {083533} (\bibinfo {year} {2025})},\ \Eprint
  {http://arxiv.org/abs/2504.16337} {arXiv:2504.16337 [astro-ph.CO]}
  \BibitemShut {NoStop}%
\bibitem [{\citenamefont {Akrami}\ \emph {et~al.}(2025)\citenamefont {Akrami},
  \citenamefont {Alestas},\ and\ \citenamefont {Nesseris}}]{Akrami:2025zlb}%
  \BibitemOpen
  \bibfield  {author} {\bibinfo {author} {\bibfnamefont {Yashar}\ \bibnamefont
  {Akrami}}, \bibinfo {author} {\bibfnamefont {George}\ \bibnamefont
  {Alestas}}, \ and\ \bibinfo {author} {\bibfnamefont {Savvas}\ \bibnamefont
  {Nesseris}},\ }\bibfield  {title} {\enquote {\bibinfo {title} {{Has DESI
  detected exponential quintessence?}}}\ }\href@noop {} {\  (\bibinfo {year}
  {2025})},\ \Eprint {http://arxiv.org/abs/2504.04226} {arXiv:2504.04226
  [astro-ph.CO]} \BibitemShut {NoStop}%
\bibitem [{\citenamefont {Bayat}\ and\ \citenamefont
  {Hertzberg}(2025)}]{Bayat:2025xfr}%
  \BibitemOpen
  \bibfield  {author} {\bibinfo {author} {\bibfnamefont {Zahra}\ \bibnamefont
  {Bayat}}\ and\ \bibinfo {author} {\bibfnamefont {Mark~P.}\ \bibnamefont
  {Hertzberg}},\ }\bibfield  {title} {\enquote {\bibinfo {title} {{Examining
  quintessence models with DESI data}},}\ }\href {\doibase
  10.1088/1475-7516/2025/08/065} {\bibfield  {journal} {\bibinfo  {journal}
  {JCAP}\ }\textbf {\bibinfo {volume} {08}},\ \bibinfo {pages} {065} (\bibinfo
  {year} {2025})},\ \Eprint {http://arxiv.org/abs/2505.18937} {arXiv:2505.18937
  [astro-ph.CO]} \BibitemShut {NoStop}%
\bibitem [{\citenamefont {Chen}\ \emph {et~al.}(2026)\citenamefont {Chen},
  \citenamefont {Cline}, \citenamefont {Muralidharan},\ and\ \citenamefont
  {Salewicz}}]{Chen:2025ywv}%
  \BibitemOpen
  \bibfield  {author} {\bibinfo {author} {\bibfnamefont {Ruiqi}\ \bibnamefont
  {Chen}}, \bibinfo {author} {\bibfnamefont {James~M.}\ \bibnamefont {Cline}},
  \bibinfo {author} {\bibfnamefont {Varun}\ \bibnamefont {Muralidharan}}, \
  and\ \bibinfo {author} {\bibfnamefont {Benjamin}\ \bibnamefont {Salewicz}},\
  }\bibfield  {title} {\enquote {\bibinfo {title} {{Quintessential dark energy
  crossing the phantom divide}},}\ }\href {\doibase
  10.1088/1475-7516/2026/03/044} {\bibfield  {journal} {\bibinfo  {journal}
  {JCAP}\ }\textbf {\bibinfo {volume} {03}},\ \bibinfo {pages} {044} (\bibinfo
  {year} {2026})},\ \Eprint {http://arxiv.org/abs/2508.19101} {arXiv:2508.19101
  [astro-ph.CO]} \BibitemShut {NoStop}%
\bibitem [{\citenamefont {Li}\ and\ \citenamefont {Zhang}(2025)}]{Li:2025ula}%
  \BibitemOpen
  \bibfield  {author} {\bibinfo {author} {\bibfnamefont {Yun-He}\ \bibnamefont
  {Li}}\ and\ \bibinfo {author} {\bibfnamefont {Xin}\ \bibnamefont {Zhang}},\
  }\bibfield  {title} {\enquote {\bibinfo {title} {{Cosmic sign-reversal:
  non-parametric reconstruction of interacting dark energy with DESI DR2}},}\
  }\href {\doibase 10.1088/1475-7516/2025/12/018} {\bibfield  {journal}
  {\bibinfo  {journal} {JCAP}\ }\textbf {\bibinfo {volume} {12}},\ \bibinfo
  {pages} {018} (\bibinfo {year} {2025})},\ \Eprint
  {http://arxiv.org/abs/2506.18477} {arXiv:2506.18477 [astro-ph.CO]}
  \BibitemShut {NoStop}%
\bibitem [{\citenamefont {{\"O}z{\"u}lker}\ \emph {et~al.}(2025)\citenamefont
  {{\"O}z{\"u}lker}, \citenamefont {Di~Valentino},\ and\ \citenamefont
  {Giar{\`e}}}]{Ozulker:2025ehg}%
  \BibitemOpen
  \bibfield  {author} {\bibinfo {author} {\bibfnamefont {Emre}\ \bibnamefont
  {{\"O}z{\"u}lker}}, \bibinfo {author} {\bibfnamefont {Eleonora}\ \bibnamefont
  {Di~Valentino}}, \ and\ \bibinfo {author} {\bibfnamefont {William}\
  \bibnamefont {Giar{\`e}}},\ }\bibfield  {title} {\enquote {\bibinfo {title}
  {{Dark Energy Crosses the Line: Quantifying and Testing the Evidence for
  Phantom Crossing}},}\ }\href@noop {} {\  (\bibinfo {year} {2025})},\ \Eprint
  {http://arxiv.org/abs/2506.19053} {arXiv:2506.19053 [astro-ph.CO]}
  \BibitemShut {NoStop}%
\bibitem [{\citenamefont {Silva}\ and\ \citenamefont
  {Nunes}(2025)}]{Silva:2025twg}%
  \BibitemOpen
  \bibfield  {author} {\bibinfo {author} {\bibfnamefont {Emanuelly}\
  \bibnamefont {Silva}}\ and\ \bibinfo {author} {\bibfnamefont {Rafael~C.}\
  \bibnamefont {Nunes}},\ }\bibfield  {title} {\enquote {\bibinfo {title}
  {{Testing signatures of phantom crossing through full-shape galaxy clustering
  analysis}},}\ }\href {\doibase 10.1088/1475-7516/2025/11/078} {\bibfield
  {journal} {\bibinfo  {journal} {JCAP}\ }\textbf {\bibinfo {volume} {11}},\
  \bibinfo {pages} {078} (\bibinfo {year} {2025})},\ \Eprint
  {http://arxiv.org/abs/2507.13989} {arXiv:2507.13989 [astro-ph.CO]}
  \BibitemShut {NoStop}%
\bibitem [{\citenamefont {Nojiri}\ \emph
  {et~al.}(2026{\natexlab{a}})\citenamefont {Nojiri}, \citenamefont
  {Odintsov},\ and\ \citenamefont {Oikonomou}}]{Nojiri:2025uew}%
  \BibitemOpen
  \bibfield  {author} {\bibinfo {author} {\bibfnamefont {Shin'ichi}\
  \bibnamefont {Nojiri}}, \bibinfo {author} {\bibfnamefont {Sergei~D.}\
  \bibnamefont {Odintsov}}, \ and\ \bibinfo {author} {\bibfnamefont {V.~K.}\
  \bibnamefont {Oikonomou}},\ }\bibfield  {title} {\enquote {\bibinfo {title}
  {{Apparent phantom crossing in Gauss{\textendash}Bonnet gravity}},}\ }\href
  {\doibase 10.1140/epjc/s10052-026-15562-x} {\bibfield  {journal} {\bibinfo
  {journal} {Eur. Phys. J. C}\ }\textbf {\bibinfo {volume} {86}},\ \bibinfo
  {pages} {353} (\bibinfo {year} {2026}{\natexlab{a}})},\ \Eprint
  {http://arxiv.org/abs/2512.06279} {arXiv:2512.06279 [gr-qc]} \BibitemShut
  {NoStop}%
\bibitem [{\citenamefont {Thanankullaphong}\ \emph {et~al.}(2026)\citenamefont
  {Thanankullaphong}, \citenamefont {Sahoo}, \citenamefont
  {Hassan~Puttasiddappa},\ and\ \citenamefont
  {Roy}}]{Thanankullaphong:2026anl}%
  \BibitemOpen
  \bibfield  {author} {\bibinfo {author} {\bibfnamefont {Phusuda}\ \bibnamefont
  {Thanankullaphong}}, \bibinfo {author} {\bibfnamefont {Prasanta}\
  \bibnamefont {Sahoo}}, \bibinfo {author} {\bibfnamefont {Prajwal}\
  \bibnamefont {Hassan~Puttasiddappa}}, \ and\ \bibinfo {author} {\bibfnamefont
  {Nandan}\ \bibnamefont {Roy}},\ }\bibfield  {title} {\enquote {\bibinfo
  {title} {{Quintom Dark Energy: Future Attractor and Phantom Crossing in Light
  of DESI DR2 Observation}},}\ }\href@noop {} {\  (\bibinfo {year} {2026})},\
  \Eprint {http://arxiv.org/abs/2601.02284} {arXiv:2601.02284 [astro-ph.CO]}
  \BibitemShut {NoStop}%
\bibitem [{\citenamefont {Gialamas}\ \emph {et~al.}(2025)\citenamefont
  {Gialamas}, \citenamefont {H{\"u}tsi}, \citenamefont {Raidal}, \citenamefont
  {Urrutia}, \citenamefont {Vasar},\ and\ \citenamefont
  {Veerm{\"a}e}}]{Gialamas:2025pwv}%
  \BibitemOpen
  \bibfield  {author} {\bibinfo {author} {\bibfnamefont {Ioannis~D.}\
  \bibnamefont {Gialamas}}, \bibinfo {author} {\bibfnamefont {Gert}\
  \bibnamefont {H{\"u}tsi}}, \bibinfo {author} {\bibfnamefont {Martti}\
  \bibnamefont {Raidal}}, \bibinfo {author} {\bibfnamefont {Juan}\ \bibnamefont
  {Urrutia}}, \bibinfo {author} {\bibfnamefont {Martin}\ \bibnamefont {Vasar}},
  \ and\ \bibinfo {author} {\bibfnamefont {Hardi}\ \bibnamefont
  {Veerm{\"a}e}},\ }\bibfield  {title} {\enquote {\bibinfo {title}
  {{Quintessence and phantoms in light of DESI 2025}},}\ }\href {\doibase
  10.1103/kdqc-y37v} {\bibfield  {journal} {\bibinfo  {journal} {Phys. Rev. D}\
  }\textbf {\bibinfo {volume} {112}},\ \bibinfo {pages} {063551} (\bibinfo
  {year} {2025})},\ \Eprint {http://arxiv.org/abs/2506.21542} {arXiv:2506.21542
  [astro-ph.CO]} \BibitemShut {NoStop}%
\bibitem [{\citenamefont {Chakraborty}\ \emph {et~al.}(2025)\citenamefont
  {Chakraborty}, \citenamefont {Chanda}, \citenamefont {Das},\ and\
  \citenamefont {Dutta}}]{Chakraborty:2025syu}%
  \BibitemOpen
  \bibfield  {author} {\bibinfo {author} {\bibfnamefont {Amlan}\ \bibnamefont
  {Chakraborty}}, \bibinfo {author} {\bibfnamefont {Prolay~K.}\ \bibnamefont
  {Chanda}}, \bibinfo {author} {\bibfnamefont {Subinoy}\ \bibnamefont {Das}}, \
  and\ \bibinfo {author} {\bibfnamefont {Koushik}\ \bibnamefont {Dutta}},\
  }\bibfield  {title} {\enquote {\bibinfo {title} {{DESI results: Hint towards
  coupled dark matter and dark energy}},}\ }\href@noop {} {\  (\bibinfo {year}
  {2025})},\ \Eprint {http://arxiv.org/abs/2503.10806} {arXiv:2503.10806
  [astro-ph.CO]} \BibitemShut {NoStop}%
\bibitem [{\citenamefont {Bedroya}\ \emph {et~al.}(2025)\citenamefont
  {Bedroya}, \citenamefont {Obied}, \citenamefont {Vafa},\ and\ \citenamefont
  {Wu}}]{Bedroya:2025fwh}%
  \BibitemOpen
  \bibfield  {author} {\bibinfo {author} {\bibfnamefont {Alek}\ \bibnamefont
  {Bedroya}}, \bibinfo {author} {\bibfnamefont {Georges}\ \bibnamefont
  {Obied}}, \bibinfo {author} {\bibfnamefont {Cumrun}\ \bibnamefont {Vafa}}, \
  and\ \bibinfo {author} {\bibfnamefont {David~H.}\ \bibnamefont {Wu}},\
  }\bibfield  {title} {\enquote {\bibinfo {title} {{Evolving Dark Sector and
  the Dark Dimension Scenario}},}\ }\href@noop {} {\  (\bibinfo {year}
  {2025})},\ \Eprint {http://arxiv.org/abs/2507.03090} {arXiv:2507.03090
  [astro-ph.CO]} \BibitemShut {NoStop}%
\bibitem [{\citenamefont {Wang}\ \emph {et~al.}(2025)\citenamefont {Wang},
  \citenamefont {Cai}, \citenamefont {Guo},\ and\ \citenamefont
  {Wang}}]{Wang:2025znm}%
  \BibitemOpen
  \bibfield  {author} {\bibinfo {author} {\bibfnamefont {Jia-Qi}\ \bibnamefont
  {Wang}}, \bibinfo {author} {\bibfnamefont {Rong-Gen}\ \bibnamefont {Cai}},
  \bibinfo {author} {\bibfnamefont {Zong-Kuan}\ \bibnamefont {Guo}}, \ and\
  \bibinfo {author} {\bibfnamefont {Shao-Jiang}\ \bibnamefont {Wang}},\
  }\bibfield  {title} {\enquote {\bibinfo {title} {{Resolving the Planck-DESI
  tension by non-minimally coupled quintessence}},}\ }\href@noop {} {\
  (\bibinfo {year} {2025})},\ \Eprint {http://arxiv.org/abs/2508.01759}
  {arXiv:2508.01759 [astro-ph.CO]} \BibitemShut {NoStop}%
\bibitem [{\citenamefont {Samanta}\ \emph {et~al.}(2025)\citenamefont
  {Samanta}, \citenamefont {Ajith},\ and\ \citenamefont
  {Panda}}]{Samanta:2025oqz}%
  \BibitemOpen
  \bibfield  {author} {\bibinfo {author} {\bibfnamefont {Atul~Ashutosh}\
  \bibnamefont {Samanta}}, \bibinfo {author} {\bibfnamefont {Abhijith}\
  \bibnamefont {Ajith}}, \ and\ \bibinfo {author} {\bibfnamefont {Sukanta}\
  \bibnamefont {Panda}},\ }\bibfield  {title} {\enquote {\bibinfo {title}
  {{Exploring Coupled Quintessence in light of CMB and DESI DR2
  measurements}},}\ }\href@noop {} {\  (\bibinfo {year} {2025})},\ \Eprint
  {http://arxiv.org/abs/2509.09624} {arXiv:2509.09624 [gr-qc]} \BibitemShut
  {NoStop}%
\bibitem [{\citenamefont {Nojiri}\ \emph {et~al.}(2025)\citenamefont {Nojiri},
  \citenamefont {Odintsov},\ and\ \citenamefont {Oikonomou}}]{Nojiri:2025low}%
  \BibitemOpen
  \bibfield  {author} {\bibinfo {author} {\bibfnamefont {Shin'ichi}\
  \bibnamefont {Nojiri}}, \bibinfo {author} {\bibfnamefont {S.~D.}\
  \bibnamefont {Odintsov}}, \ and\ \bibinfo {author} {\bibfnamefont {V.~K.}\
  \bibnamefont {Oikonomou}},\ }\bibfield  {title} {\enquote {\bibinfo {title}
  {{Phantom crossing and oscillating dark energy with F(R) gravity}},}\ }\href
  {\doibase 10.1103/16yg-966k} {\bibfield  {journal} {\bibinfo  {journal}
  {Phys. Rev. D}\ }\textbf {\bibinfo {volume} {112}},\ \bibinfo {pages}
  {104035} (\bibinfo {year} {2025})},\ \Eprint
  {http://arxiv.org/abs/2506.21010} {arXiv:2506.21010 [gr-qc]} \BibitemShut
  {NoStop}%
\bibitem [{\citenamefont {S{\'a}nchez~L{\'o}pez}\ \emph
  {et~al.}(2025)\citenamefont {S{\'a}nchez~L{\'o}pez}, \citenamefont {Karam},\
  and\ \citenamefont {Hazra}}]{SanchezLopez:2025uzw}%
  \BibitemOpen
  \bibfield  {author} {\bibinfo {author} {\bibfnamefont {Samuel}\ \bibnamefont
  {S{\'a}nchez~L{\'o}pez}}, \bibinfo {author} {\bibfnamefont {Alexandros}\
  \bibnamefont {Karam}}, \ and\ \bibinfo {author} {\bibfnamefont
  {Dhiraj~Kumar}\ \bibnamefont {Hazra}},\ }\bibfield  {title} {\enquote
  {\bibinfo {title} {{Non-Minimally Coupled Quintessence in Light of DESI}},}\
  }\href@noop {} {\  (\bibinfo {year} {2025})},\ \Eprint
  {http://arxiv.org/abs/2510.14941} {arXiv:2510.14941 [astro-ph.CO]}
  \BibitemShut {NoStop}%
\bibitem [{\citenamefont {La~Penna}\ \emph {et~al.}(2026)\citenamefont
  {La~Penna}, \citenamefont {Notari},\ and\ \citenamefont
  {Redi}}]{LaPenna:2026avs}%
  \BibitemOpen
  \bibfield  {author} {\bibinfo {author} {\bibfnamefont {Lorenzo}\ \bibnamefont
  {La~Penna}}, \bibinfo {author} {\bibfnamefont {Alessio}\ \bibnamefont
  {Notari}}, \ and\ \bibinfo {author} {\bibfnamefont {Michele}\ \bibnamefont
  {Redi}},\ }\bibfield  {title} {\enquote {\bibinfo {title} {{Mimicking Phantom
  Dark Energy with Evolving Dark Matter Mass}},}\ }\href@noop {} {\  (\bibinfo
  {year} {2026})},\ \Eprint {http://arxiv.org/abs/2601.05235} {arXiv:2601.05235
  [astro-ph.CO]} \BibitemShut {NoStop}%
\bibitem [{\citenamefont {Li}\ \emph {et~al.}(2026{\natexlab{b}})\citenamefont
  {Li}, \citenamefont {Giar{\`e}}, \citenamefont {Du}, \citenamefont {Li},
  \citenamefont {Di~Valentino}, \citenamefont {Zhang},\ and\ \citenamefont
  {Zhang}}]{Li:2026xaz}%
  \BibitemOpen
  \bibfield  {author} {\bibinfo {author} {\bibfnamefont {Tian-Nuo}\
  \bibnamefont {Li}}, \bibinfo {author} {\bibfnamefont {William}\ \bibnamefont
  {Giar{\`e}}}, \bibinfo {author} {\bibfnamefont {Guo-Hong}\ \bibnamefont
  {Du}}, \bibinfo {author} {\bibfnamefont {Yun-He}\ \bibnamefont {Li}},
  \bibinfo {author} {\bibfnamefont {Eleonora}\ \bibnamefont {Di~Valentino}},
  \bibinfo {author} {\bibfnamefont {Jing-Fei}\ \bibnamefont {Zhang}}, \ and\
  \bibinfo {author} {\bibfnamefont {Xin}\ \bibnamefont {Zhang}},\ }\bibfield
  {title} {\enquote {\bibinfo {title} {{Strong Evidence for Dark Sector
  Interactions}},}\ }\href@noop {} {\  (\bibinfo {year}
  {2026}{\natexlab{b}})},\ \Eprint {http://arxiv.org/abs/2601.07361}
  {arXiv:2601.07361 [astro-ph.CO]} \BibitemShut {NoStop}%
\bibitem [{\citenamefont {Odintsov}\ and\ \citenamefont
  {Oikonomou}(2026)}]{Odintsov:2026doe}%
  \BibitemOpen
  \bibfield  {author} {\bibinfo {author} {\bibfnamefont {S.~D.}\ \bibnamefont
  {Odintsov}}\ and\ \bibinfo {author} {\bibfnamefont {V.~K.}\ \bibnamefont
  {Oikonomou}},\ }\bibfield  {title} {\enquote {\bibinfo {title} {{R2-corrected
  Tachyon Scalar Field Inflation, the ACT Data, and Phantom Transition}},}\
  }\href {\doibase 10.1016/j.nuclphysb.2026.117384} {\bibfield  {journal}
  {\bibinfo  {journal} {Nucl. Phys. B}\ }\textbf {\bibinfo {volume} {1025}},\
  \bibinfo {pages} {117384} (\bibinfo {year} {2026})},\ \Eprint
  {http://arxiv.org/abs/2601.21364} {arXiv:2601.21364 [gr-qc]} \BibitemShut
  {NoStop}%
\bibitem [{\citenamefont {Nojiri}\ \emph
  {et~al.}(2026{\natexlab{b}})\citenamefont {Nojiri}, \citenamefont
  {Odintsov},\ and\ \citenamefont {Oikonomou}}]{Nojiri:2026uvn}%
  \BibitemOpen
  \bibfield  {author} {\bibinfo {author} {\bibfnamefont {Shin'ichi}\
  \bibnamefont {Nojiri}}, \bibinfo {author} {\bibfnamefont {S.~D.}\
  \bibnamefont {Odintsov}}, \ and\ \bibinfo {author} {\bibfnamefont {V.~K.}\
  \bibnamefont {Oikonomou}},\ }\bibfield  {title} {\enquote {\bibinfo {title}
  {{Is Phantom Divide Crossing in General Relativity Completely Impossible?
  Shortcomings and Possible Solutions}},}\ }\href@noop {} {\  (\bibinfo {year}
  {2026}{\natexlab{b}})},\ \Eprint {http://arxiv.org/abs/2601.21356}
  {arXiv:2601.21356 [gr-qc]} \BibitemShut {NoStop}%
\bibitem [{\citenamefont {Roy~Choudhury}\ and\ \citenamefont
  {Okumura}(2024)}]{RoyChoudhury:2024wri}%
  \BibitemOpen
  \bibfield  {author} {\bibinfo {author} {\bibfnamefont {Shouvik}\ \bibnamefont
  {Roy~Choudhury}}\ and\ \bibinfo {author} {\bibfnamefont {Teppei}\
  \bibnamefont {Okumura}},\ }\bibfield  {title} {\enquote {\bibinfo {title}
  {{Updated Cosmological Constraints in Extended Parameter Space with Planck
  PR4, DESI Baryon Acoustic Oscillations, and Supernovae: Dynamical Dark
  Energy, Neutrino Masses, Lensing Anomaly, and the Hubble Tension}},}\ }\href
  {\doibase 10.3847/2041-8213/ad8c26} {\bibfield  {journal} {\bibinfo
  {journal} {Astrophys. J. Lett.}\ }\textbf {\bibinfo {volume} {976}},\
  \bibinfo {pages} {L11} (\bibinfo {year} {2024})},\ \Eprint
  {http://arxiv.org/abs/2409.13022} {arXiv:2409.13022 [astro-ph.CO]}
  \BibitemShut {NoStop}%
\bibitem [{\citenamefont {Roy~Choudhury}(2025)}]{RoyChoudhury:2025dhe}%
  \BibitemOpen
  \bibfield  {author} {\bibinfo {author} {\bibfnamefont {Shouvik}\ \bibnamefont
  {Roy~Choudhury}},\ }\bibfield  {title} {\enquote {\bibinfo {title}
  {{Cosmology in Extended Parameter Space with DESI Data Release 2 Baryon
  Acoustic Oscillations: A 2{\ensuremath{\sigma}}+ Detection of Nonzero
  Neutrino Masses with an Update on Dynamical Dark Energy and Lensing
  Anomaly}},}\ }\href {\doibase 10.3847/2041-8213/ade1cc} {\bibfield  {journal}
  {\bibinfo  {journal} {Astrophys. J. Lett.}\ }\textbf {\bibinfo {volume}
  {986}},\ \bibinfo {pages} {L31} (\bibinfo {year} {2025})},\ \Eprint
  {http://arxiv.org/abs/2504.15340} {arXiv:2504.15340 [astro-ph.CO]}
  \BibitemShut {NoStop}%
\bibitem [{\citenamefont {Ong}\ \emph {et~al.}(2026)\citenamefont {Ong},
  \citenamefont {Yallup},\ and\ \citenamefont {Handley}}]{Ong:2026tta}%
  \BibitemOpen
  \bibfield  {author} {\bibinfo {author} {\bibfnamefont {Dily Duan~Yi}\
  \bibnamefont {Ong}}, \bibinfo {author} {\bibfnamefont {David}\ \bibnamefont
  {Yallup}}, \ and\ \bibinfo {author} {\bibfnamefont {Will}\ \bibnamefont
  {Handley}},\ }\bibfield  {title} {\enquote {\bibinfo {title} {{The Bayesian
  view of DESI DR2: Evidence and tension in a combined analysis with CMB and
  supernovae across cosmological models}},}\ }\href@noop {} {\  (\bibinfo
  {year} {2026})},\ \Eprint {http://arxiv.org/abs/2603.05472} {arXiv:2603.05472
  [astro-ph.CO]} \BibitemShut {NoStop}%
\bibitem [{\citenamefont {Wang}\ and\ \citenamefont
  {Mota}(2025)}]{Wang:2025bkk}%
  \BibitemOpen
  \bibfield  {author} {\bibinfo {author} {\bibfnamefont {Deng}\ \bibnamefont
  {Wang}}\ and\ \bibinfo {author} {\bibfnamefont {David}\ \bibnamefont
  {Mota}},\ }\bibfield  {title} {\enquote {\bibinfo {title} {{Did DESI DR2
  truly reveal dynamical dark energy?}}}\ }\href {\doibase
  10.1140/epjc/s10052-025-15076-y} {\bibfield  {journal} {\bibinfo  {journal}
  {Eur. Phys. J. C}\ }\textbf {\bibinfo {volume} {85}},\ \bibinfo {pages}
  {1356} (\bibinfo {year} {2025})},\ \Eprint {http://arxiv.org/abs/2504.15222}
  {arXiv:2504.15222 [astro-ph.CO]} \BibitemShut {NoStop}%
\bibitem [{\citenamefont {Roy~Choudhury}\ \emph {et~al.}(2025)\citenamefont
  {Roy~Choudhury}, \citenamefont {Okumura},\ and\ \citenamefont
  {Umetsu}}]{RoyChoudhury:2025iis}%
  \BibitemOpen
  \bibfield  {author} {\bibinfo {author} {\bibfnamefont {Shouvik}\ \bibnamefont
  {Roy~Choudhury}}, \bibinfo {author} {\bibfnamefont {Teppei}\ \bibnamefont
  {Okumura}}, \ and\ \bibinfo {author} {\bibfnamefont {Keiichi}\ \bibnamefont
  {Umetsu}},\ }\bibfield  {title} {\enquote {\bibinfo {title} {{Cosmological
  Constraints on Nonphantom Dynamical Dark Energy with DESI Data Release 2
  Baryon Acoustic Oscillations: A 3{\ensuremath{\sigma}}+ Lensing Anomaly}},}\
  }\href {\doibase 10.3847/2041-8213/ae1a64} {\bibfield  {journal} {\bibinfo
  {journal} {Astrophys. J. Lett.}\ }\textbf {\bibinfo {volume} {994}},\
  \bibinfo {pages} {L26} (\bibinfo {year} {2025})},\ \Eprint
  {http://arxiv.org/abs/2509.26144} {arXiv:2509.26144 [astro-ph.CO]}
  \BibitemShut {NoStop}%
\bibitem [{\citenamefont {Cheng}\ \emph {et~al.}(2026)\citenamefont {Cheng},
  \citenamefont {Pan},\ and\ \citenamefont {Di~Valentino}}]{Cheng:2025yue}%
  \BibitemOpen
  \bibfield  {author} {\bibinfo {author} {\bibfnamefont {Hanyu}\ \bibnamefont
  {Cheng}}, \bibinfo {author} {\bibfnamefont {Supriya}\ \bibnamefont {Pan}}, \
  and\ \bibinfo {author} {\bibfnamefont {Eleonora}\ \bibnamefont
  {Di~Valentino}},\ }\bibfield  {title} {\enquote {\bibinfo {title} {{Beyond
  Two Parameters: Revisiting Dark Energy with the Latest Cosmic Probes}},}\
  }\href {\doibase 10.3847/1538-4357/ae3a8f} {\bibfield  {journal} {\bibinfo
  {journal} {Astrophys. J.}\ }\textbf {\bibinfo {volume} {999}},\ \bibinfo
  {pages} {190} (\bibinfo {year} {2026})},\ \Eprint
  {http://arxiv.org/abs/2512.09866} {arXiv:2512.09866 [astro-ph.CO]}
  \BibitemShut {NoStop}%
\bibitem [{\citenamefont {Colg{\'a}in}\ \emph {et~al.}(2025)\citenamefont
  {Colg{\'a}in}, \citenamefont {Pourojaghi},\ and\ \citenamefont
  {Sheikh-Jabbari}}]{Colgain:2025fct}%
  \BibitemOpen
  \bibfield  {author} {\bibinfo {author} {\bibfnamefont {Eoin~{\'O}.}\
  \bibnamefont {Colg{\'a}in}}, \bibinfo {author} {\bibfnamefont {Saeed}\
  \bibnamefont {Pourojaghi}}, \ and\ \bibinfo {author} {\bibfnamefont {M.~M.}\
  \bibnamefont {Sheikh-Jabbari}},\ }\bibfield  {title} {\enquote {\bibinfo
  {title} {{On the Analysis Dependence of DESI Dynamical Dark Energy}},}\
  }\href {\doibase 10.3390/galaxies13060133} {\bibfield  {journal} {\bibinfo
  {journal} {Galaxies}\ }\textbf {\bibinfo {volume} {13}},\ \bibinfo {pages}
  {133} (\bibinfo {year} {2025})},\ \Eprint {http://arxiv.org/abs/2505.19029}
  {arXiv:2505.19029 [astro-ph.CO]} \BibitemShut {NoStop}%
\bibitem [{\citenamefont {Costa}\ \emph {et~al.}(2017)\citenamefont {Costa},
  \citenamefont {Xu}, \citenamefont {Wang},\ and\ \citenamefont
  {Abdalla}}]{Costa:2016tpb}%
  \BibitemOpen
  \bibfield  {author} {\bibinfo {author} {\bibfnamefont {Andr{\'e}~A.}\
  \bibnamefont {Costa}}, \bibinfo {author} {\bibfnamefont {Xiao-Dong}\
  \bibnamefont {Xu}}, \bibinfo {author} {\bibfnamefont {Bin}\ \bibnamefont
  {Wang}}, \ and\ \bibinfo {author} {\bibfnamefont {E.}~\bibnamefont
  {Abdalla}},\ }\bibfield  {title} {\enquote {\bibinfo {title} {{Constraints on
  interacting dark energy models from Planck 2015 and redshift-space distortion
  data}},}\ }\href {\doibase 10.1088/1475-7516/2017/01/028} {\bibfield
  {journal} {\bibinfo  {journal} {JCAP}\ }\textbf {\bibinfo {volume} {01}},\
  \bibinfo {pages} {028} (\bibinfo {year} {2017})},\ \Eprint
  {http://arxiv.org/abs/1605.04138} {arXiv:1605.04138 [astro-ph.CO]}
  \BibitemShut {NoStop}%
\bibitem [{\citenamefont {Ade}\ \emph {et~al.}(2016)\citenamefont {Ade} \emph
  {et~al.}}]{Planck:2015bue}%
  \BibitemOpen
  \bibfield  {author} {\bibinfo {author} {\bibfnamefont {P.~A.~R.}\
  \bibnamefont {Ade}} \emph {et~al.} (\bibinfo {collaboration} {Planck}),\
  }\bibfield  {title} {\enquote {\bibinfo {title} {{Planck 2015 results. XIV.
  Dark energy and modified gravity}},}\ }\href {\doibase
  10.1051/0004-6361/201525814} {\bibfield  {journal} {\bibinfo  {journal}
  {Astron. Astrophys.}\ }\textbf {\bibinfo {volume} {594}},\ \bibinfo {pages}
  {A14} (\bibinfo {year} {2016})},\ \Eprint {http://arxiv.org/abs/1502.01590}
  {arXiv:1502.01590 [astro-ph.CO]} \BibitemShut {NoStop}%
\bibitem [{\citenamefont {Wang}\ \emph {et~al.}(2026)\citenamefont {Wang},
  \citenamefont {Cai}, \citenamefont {Guo}, \citenamefont {Li}, \citenamefont
  {Wang},\ and\ \citenamefont {Zhang}}]{Wang:2026wrk}%
  \BibitemOpen
  \bibfield  {author} {\bibinfo {author} {\bibfnamefont {Jia-Qi}\ \bibnamefont
  {Wang}}, \bibinfo {author} {\bibfnamefont {Rong-Gen}\ \bibnamefont {Cai}},
  \bibinfo {author} {\bibfnamefont {Zong-Kuan}\ \bibnamefont {Guo}}, \bibinfo
  {author} {\bibfnamefont {Yun-He}\ \bibnamefont {Li}}, \bibinfo {author}
  {\bibfnamefont {Shao-Jiang}\ \bibnamefont {Wang}}, \ and\ \bibinfo {author}
  {\bibfnamefont {Xin}\ \bibnamefont {Zhang}},\ }\bibfield  {title} {\enquote
  {\bibinfo {title} {{Non-minimally coupled quintessence with sign-switching
  interaction}},}\ }\href@noop {} {\  (\bibinfo {year} {2026})},\ \Eprint
  {http://arxiv.org/abs/2604.02204} {arXiv:2604.02204 [astro-ph.CO]}
  \BibitemShut {NoStop}%
\end{thebibliography}%

\end{document}